\documentclass[journal]{IEEEtran}
\usepackage[noadjust]{cite}
\usepackage[english]{babel}
\usepackage{graphicx}
\usepackage{tikz}
\usepackage{circuitikz}
\usepackage{xcolor}
\usepackage{picinpar}
\usepackage{amsmath}
\usepackage{amsthm}
\usepackage{mathrsfs}
\usepackage[T1]{fontenc}
\usepackage{url}
\usepackage{inputenc}
\usepackage[linesnumbered,ruled,vlined]{algorithm2e}
\usepackage{colortbl}
\usepackage{soul}
\usepackage{multirow}
\usepackage{pifont}
\usepackage{color}
\usepackage{alltt}
\usepackage[hidelinks]{hyperref}
\usepackage{enumerate}
\usepackage{tabularx}
\usepackage{tcolorbox}
\usepackage{eso-pic}
\usepackage{siunitx}
\usepackage{makecell}
\usepackage{epstopdf}
\usepackage{subcaption}
\usepackage{lipsum}
\usepackage{array}
\usepackage{stfloats}
\usepackage{pbox}
\usepackage{float}
\usepackage{ragged2e}
\usepackage{amssymb}

\newtheoremstyle{semicolon}
  {3pt}{3pt}
  {\itshape}{}
  {\bfseries}
  {:}
  { }
  {\thmname{#1}\ \thmnumber{#2}\ \thmnote{ {\normalfont(#3)}}}

\theoremstyle{semicolon}
\newtheorem{theorem}{Theorem}

\usepackage[english]{babel}
\makeatletter
\renewcommand{\fnum@figure}{Fig. \thefigure}
\makeatother
\renewcommand{\qedsymbol}{$\blacksquare$}

\begin{document}
 \title{Gaussian Mixture Model Based Bayesian Learning for Sparse Channel Estimation in Orthogonal Time Frequency Space Modulated Systems}

\author{
	\vskip 1em
	{ Surbhi Gehlot, \emph{Graduate Student Member, IEEE},
Suraj Srivastava, \emph{Member, IEEE},
Sandeep Kumar Yadav, \emph{Member, IEEE},
and Lajos Hanzo, \emph{Life Fellow, IEEE}
	}
   \thanks {\par Surbhi Gehlot, Suraj Srivastava and Sandeep Kumar Yadav are with the Department of Electrical Engineering, Indian Institute of Technology, Jodhpur, India 342030 (e-mail: gehlot.5@iitj.ac.in, surajsri@iitj.ac.in, sy@iitj.ac.in).
   \par Lajos Hanzo is with the Department of Electronics and Computer Science,
University of Southampton, SO17 1BJ Southampton, United Kingdom (email:
lh@ecs.soton.ac.uk).
\par The work of S. Srivastava was supported in part by IIT Jodhpur's Research Grant No. I/RIG/SUS/20240043; in part by Anusandhan National Research Foundation's PM-ECRG/2024/478/ENS and ANRF/ARG/2025/005895/ENS; and in part by Telecom Technology Development Fund (TTDF) under Grant TTDF/6G/368. Lajos Hanzo would like to gratefully acknowledge the financial support of the following Engineering and Physical Sciences Research Council (EPSRC) projects: Platform for Driving Ultimate Connectivity (TITAN) (EP/X04047X/1; EP/Y037243/1); Robust and Reliable Quantum Computing (RoaRQ, EP/W032635/1); PerCom (EP/X012301/1). S. Srivastava and L. Hanzo jointly acknowledge the funding support provided to ICON-project by DST and UKRI-EPSRC under India-UK Joint opportunity in Telecommunications Research.
   }
    }
\maketitle

\AddToShipoutPictureFG*{%
  \AtPageLowerLeft{%
    \ifnum\value{page}=1
    \hspace*{0.05\textwidth}%
    \raisebox{0.2cm}{%
          \begin{tcolorbox}[colback=gray!10, colframe=black, width=0.9\textwidth, boxrule=0.5pt]
      \footnotesize
      This paper has been accepted for publication in the IEEE Open Journal of Vehicular Technology.
      The authors' version is available at IEEE Xplore: \texttt{10.1109/OJVT.2026.3676870}.
      \end{tcolorbox}
    }%
    \fi
  }%
}

\begin{abstract}
A novel Gaussian mixture model (GMM)–aided sparse Bayesian learning (SBL) framework is proposed for channel state information (CSI) estimation in orthogonal time-frequency space (OTFS) modulated systems. The key attribute of the proposed algorithm lies in casting CSI recovery as an SBL inference problem, where posterior distributions are iteratively refined under a hierarchical GMM prior. Using this approach, the sparsity-inducing variances beneficially promote sparsity in the delay–Doppler (DD) domain, while additionally augmenting the capability of SBL to exploit channel statistics more effectively. 
Moreover, to fully exploit the GMM’s ability to approximate arbitrary probability density functions and model complex multipath fading scenarios, the channel statistics are represented using a complex Gaussian mixture. Simultaneously, the method leverages time-domain (TD) pilots without requiring wasteful  DD domain guard intervals, thereby ensuring low pilot overhead and high spectral efficiency. The CSI recovered is subsequently applied in a linear minimum mean square error (MMSE) detector for reliable data detection. To benchmark performance, the Oracle-MMSE and the Bayesian Cram\'{e}r-Rao lower bound (BCRLB) are also derived. Our simulation results demonstrate significant performance improvement over the state-of-the-art sparse estimation methods.
\end{abstract}
\section{INTRODUCTION}
Next-generation (NG) wireless networks are expected to deliver ultra-high data rates in highly dynamic environments. Typical scenarios include high-speed rail, vehicular, and aerial communications, where extreme user mobility induces significant delay spread from multipath propagation and severe Doppler shifts from high relative velocity \cite{zhang2019aeronautical,liu2019recent}. Notably, under such doubly selective channel conditions, the performance of the ubiquitous orthogonal frequency-division multiplexing (OFDM) severely degrades again primarily due to the grave inter-carrier interference induced by Doppler shifts at high mobility and carrier frequencies \cite{wu2016survey}. To address this limitation, the Doppler-resilient orthogonal time-frequency space (OTFS) modulation proposed by Hadani \textit{et al.} has emerged as a promising alternative candidate \cite{hadani2017orthogonal}. By mapping information symbols onto the delay–Doppler (DD) domain instead of the conventional time–frequency (TF) domain, OTFS offers a channel representation that remains nearly invariant under high mobility \cite{raviteja2019embedded,monk2016otfs,mohammed2021derivation}. However, the performance gains of OTFS critically depend on the availability of highly accurate  DD domain channel state information (CSI) \cite{srivastava2021bayesian}. Consequently, as discussed next, numerous studies have investigated efficient CSI estimation methods designed for OTFS systems.

\subsection{Literature Review}
Channel estimation in OTFS systems has been widely explored, with the earliest approaches belonging to impulse-based CSI estimation. Representative treatises in this category are by Hadani and Monk \cite{hadani2018otfs}, as well as the solutions in \cite{raviteja2018interference, ramachandran2020otfs}, which developed an end-to-end  DD domain input–output relationship expressed as a two-dimensional circular convolution between the  DD domain signal and the  DD domain channel. While these designs were conceptually simple, their major drawback was the need for a full frame of pilots, leading to excessive pilot overhead and severely reduced spectral efficiency. To alleviate this shortcoming, Raviteja \textit{et al.} \cite{raviteja2019embedded} proposed a threshold-based CSI estimator and low complexity data detector within the same frame. The method embedded a known high-power pilot symbol at a chosen DD coordinate with guard bands around the pilot to isolate its interference from the data. However, it typically requires high transmit power, which raises the peak-to-average power ratio (PAPR), and large guard regions in the DD grid that reduce spectral efficiency; additionally, its threshold-based detection is signal-to-noise ratio (SNR) sensitive and requires careful tuning. 

The subsequent literature of OTFS has increasingly focused on exploiting the intrinsic sparsity of wireless channels in the  DD domain. Leveraging this sparse representation not only provides a structurally informed estimation framework but also helps reduce pilot overhead and improve accuracy compared to conventional pilot-based schemes. Building on this idea, in \cite{srivastava2022delay, shen2019channel} conventional compressed sensing techniques such as orthogonal matching pursuit (OMP) have been refined for exploiting the associated structured multi-dimensional sparsity, jointly capturing the sparsity in delay, Doppler and angular domains. Beyond greedy compressed sensing (CS) techniques, a parallel line of work has developed Bayesian learning (BL) based estimators for CSI estimation in OTFS systems. Zhao \textit{et al.} \cite{zhao2020sparse} formulated a sparse Bayesian learning (SBL) aided framework for sparse channel estimation in OTFS systems. They suggested a novel  DD domain pilot pattern that eliminated the need for guard symbols and maintained equal power for pilots and data, thereby reducing both the PAPR and pilot overhead. However, a fundamental limitation of the  DD domain piloting strategy is that the pilot arrangement must be carefully designed to avoid interference with data symbols, which can complicate the frame structure. Addressing this limitation, Srivastava \textit{et al.} \cite{srivastava2021bayesian} introduced a more flexible TF domain piloting strategy. The SBL framework using this pilot design significantly reduced the pilot overhead, while approaching the Bayesian Cram\'{e}r-Rao lower bound (BCRLB). Building upon these foundations, subsequently an off-grid SBL approach was developed in \cite{wei2022off}. This method formulates channel estimation as a 1D off-grid sparse signal recovery problem within the SBL framework. It effectively separates the estimation of on-grid and off-grid delay/Doppler components, modeling the latter as hyperparameters to be estimated via expectation–maximization (EM). A group-sparse Bayesian learning framework was developed for exploiting structural sparsity in the  DD domain \cite{srivastava2021bayesian1}. The model effectively captures the sparsity profile across DD-grid bins, significantly enhancing estimation accuracy even in pilot-limited OTFS regimes. As a further advancement a parametric estimation-based algorithm was proposed in \cite{khan2021low}. This proposal goes beyond sparse recovery algorithms for CSI in OTFS by advocating a modified maximum likelihood estimator and a two-step estimator, which rely on fine DD resolution to decouple the joint estimation of channel gains into independent estimation tasks using the delay and Doppler bins of each path.

\begin{table*}[!t]
\centering
\caption{Contrasting Key Features of Proposed and Existing OTFS CSI Estimation Methods (R = Required, NR = Not Required)}
\label{tab:otfs_final_grid}
\scriptsize
\setlength{\tabcolsep}{3.5pt}
\renewcommand{\arraystretch}{1.05}
\resizebox{\textwidth}{!}{%
\begin{tabular}{|p{3.3cm}|*{13}{c|}}
\hline
 & \cite{raviteja2018interference} & \cite{raviteja2019embedded} & \cite{srivastava2022delay} & \cite{zhao2020sparse} & \cite{srivastava2021bayesian} & \cite{srivastava2021bayesian1} & \cite{khan2021low} & \cite{li2022residual} & \cite{guo2024otfs} & \cite{zhang2024sparse}
 & \cite{payami2025sparse} & \cite{mattu2024learning} & Proposed \\
\hline

 DD domain sparsity &  &  & \ding{51} & \ding{51} & \ding{51} & \ding{51} & \ding{51} & \ding{51} & \ding{51} & \ding{51} &  & \ding{51} & \ding{51} \\
\hline
CSI labels & NR & NR & NR & NR & NR & NR & NR & R & R & R & R & R & NR \\
\hline
Flexible pilot overhead &  &  & \ding{51} &  & \ding{51} & \ding{51} & \ding{51} &  & \ding{51} &  & \ding{51} & \ding{51} & \ding{51} \\
\hline
BCRLB &  &  & \ding{51} & \ding{51} & \ding{51} & \ding{51} &  &  &  &  &  &  & \ding{51} \\
\hline
DD-guard & R & R & NR & R & NR & NR & NR & R & R & R & R & R & NR \\
\hline
Rectangular pulse shaping &  &  & \ding{51} &  & \ding{51} &  & \ding{51} & &  &  & \ding{51} &  & \ding{51} \\
\hline
Offline training & NR & NR & NR & NR & NR & NR & NR & R & R & R & R & R & NR \\
\hline
Multiple training snapshots &  &  &  &  & \ding{51} &  &  &  &  &  &  &  & \ding{51} \\
\hline
\textbf{GMM channel modeling} &  &  &  &  &  &  &  &  &  &  &  &  & \ding{51} \\
\hline
\textbf{Sparsity under GMM prior} &  &  &  &  &  &  &  &  &  &  &  &  & \ding{51} \\
\hline
\textbf{Unified GMM--SBL framework} &  &  &  &  &  &  &  &  &  &  &  &  & \ding{51} \\
\hline
\end{tabular}
}
\end{table*}

Recent years have also witnessed the emergence of deep-learning (DL) techniques for channel estimation in OTFS systems. A number of these treatises \cite{li2022residual, guo2024otfs, zhang2024sparse} adopt a hybrid estimate-and-denoise paradigm, where a conventional estimator produces a coarse DD CSI estimator that is then refined by a neural network. Furthermore, several papers have conceived alternative DL based learning techniques for OTFS channel estimation. Payami \textit{et al.} \cite{payami2025sparse} introduced convolutional neural network (CNN) based support-learning architectures that directly infer the active DD support without an explicit sensing matrix. Their approach first identifies the support locations of the dominant channel taps and then estimates the corresponding amplitudes. A supervised neural network is proposed in \cite{mattu2024learning} to learn the mapping from received TF-domain pilots to the underlying  DD domain channel, whereas a long short-term memory (LSTM) based model in \cite{dos2024lstm} leverages temporal correlations across pilot sequences to learn the underlying  DD domain channel dynamics. Despite the notable estimation accuracy achieved by DL-based approaches, their employment remains constrained by the need for extensive labelled training datasets and frequent retraining.  Moreover, their black-box nature reduces interpretability, motivating the development of robust, model-driven estimation frameworks over purely data-driven ones.

Existing research gaps in CSI estimation methods for OTFS systems as discussed in Table \ref{tab:otfs_final_grid} underscore the need for a more efficient, interpretable, and statistically robust  DD domain estimation framework. Towards this end, we propose a novel Gaussian mixture model–aided sparse Bayesian learning (GMM-SBL) framework, whose key contributions are articulated in the following section.

\subsection{Contributions of the Paper}
\begin{enumerate}

    \item To enhance the capability of conventional SBL, a novel GMM-SBL framework is proposed for OTFS channel estimation. Conventional SBL relies on a single Gaussian prior, which limits its expressiveness in modelling clustered sparsity as observed in practical wireless channels. To address this, the proposed learning strategy preserves the fundamental sparsity-promoting mechanism of SBL in the DD domain, while enhancing the statistical modelling flexibility through Gaussian mixture–based priors. The approach is supported by a theoretical analysis of sparsity guarantees under mixture priors and it is implemented via a unified EM algorithm that jointly updates the adaptive mixture weights and variance estimates directly from the received pilot observations.
    
    \medskip

    \item By representing the channel distribution as a weighted combination of Gaussian components, the prior harnesses the full potential of the proposed GMM-SBL method. Furthermore, the learned mixture weights and component covariances strengthen the capability of SBL to exploit channel statistics beyond the restrictive zero-mean complex Gaussian assumption, offering a more practical and flexible representation of multipath channels.
     \medskip
    \item By transmitting pilots in the time domain, the framework enables flexible pilot placement without requiring precise DD grid locations. This facilitates full utilisation of the DD grid for data transmission, thereby substantially reducing pilot overhead and improving spectral efficiency. Following the sparse channel estimation, the recovered CSI is incorporated into a linear minimum mean square error (MMSE) detector, which leverages the full posterior statistics for reliable symbol detection.
      \medskip
    
    \item To establish performance references, an Oracle-based MMSE and the BCRLB of GMM priors are derived. Furthermore, the robustness of the proposed algorithm
is validated through extensive simulations under varying mixture weights, cluster means, number of clusters, pilot overhead and training sample sizes, consistently demonstrating superior performance over state-of-the-art sparse estimation methods.

\end{enumerate}
The remainder of this paper is organised as follows. Section \ref{OTFS System Description} introduces the OTFS system model. Section \ref{Spare CSI Estimation Model for OTFS Systems} develops the proposed GMM–SBL based sparse CSI estimation framework. Section \ref{Performance Benchmark} describes the performance benchmarks, and Section \ref{Results and Discussion} presents extensive simulation results, comparisons with existing methods, and the computational complexity assessment of the proposed algorithm. Finally, Section \ref{Conclusion} concludes the paper.

\medskip
\textit{Notation} - Boldface lowercase and uppercase letters denote column vectors and matrices, respectively.  
The vectorization of a matrix $\mathbf{A}$ is denoted by $\mathrm{vec}(\mathbf{A})$, while $\mathrm{vec}^{-1}(\mathbf{a})$ denotes the inverse operation that reconstructs the original matrix.  
A standard identity of the vectorization operator $\mathrm{vec}(\mathbf{A}\mathbf{B}\mathbf{C}) = (\mathbf{C}^T \otimes \mathbf{A}) \, \mathrm{vec}(\mathbf{B})$, is used in the paper, where $\otimes$ denotes the Kronecker product. The Hermitian transpose of a matrix $\mathbf{A}$ is denoted by $\mathbf{A}^H$, and the expectation operator by $\mathbb{E}[\cdot]$.

\begin{figure*}[t]
\centerline{\includegraphics[width=\textwidth]{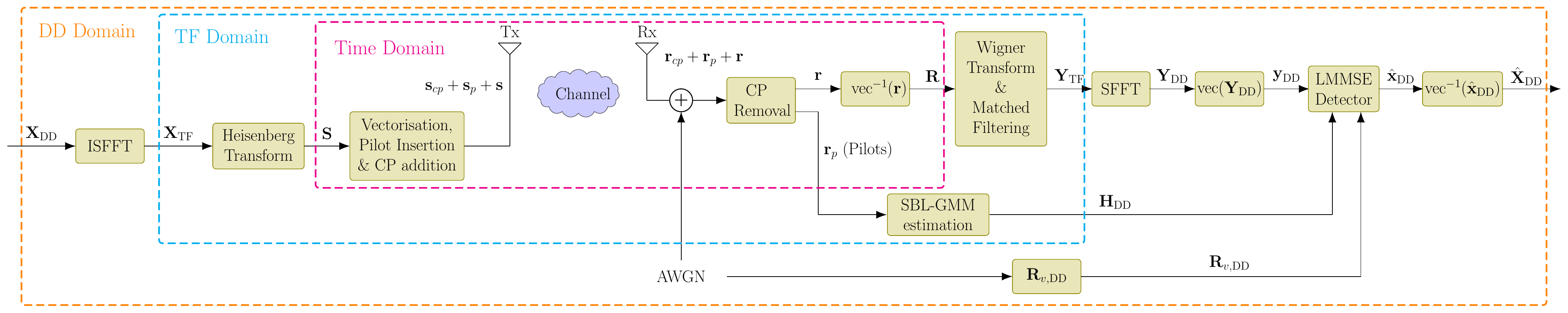}}
 \caption{Schematic diagram of GMM-SBL scheme for CP-aided OTFS transceiver.}
\label{OTFS_Transceiver}
\end{figure*}

\section{OTFS System Description}
\label{OTFS System Description}
OTFS is a $2$D modulation technique in which information symbols are mapped in the  DD domain (Fig. \ref{OTFS_Transceiver}). Consider a  DD domain grid 
and the corresponding time frequency (TF) grid.  
Let the OTFS frame duration and the bandwidth occupied be \(T_f = N T\) and \(B = M\Delta f\), respectively, where \(N\) and \(M\) denote the number of time and frequency samples on the TF-grid. Furthermore, \(T\) (seconds) denotes the symbol interval, while \(\Delta f\) (Hz) is the subcarrier spacing, obeying \(T\Delta f = 1\). Next, the signal processing operations underlying OTFS are described.

\subsection{OTFS Modulation}
\label{subsec:modulation}

Let \(\mathbf{X}_{\mathrm{DD}} \in \mathbb{C}^{M \times N}\) denote the matrix of  DD domain information symbols, where \(\mathbf{X}_{\mathrm{DD}}(\ell, c)\) is placed at delay index \(\ell\) and Doppler index \(c\). Let \(\mathbf{X}_{\mathrm{TF}}(m, n)\) be the TF-domain symbol transmitted on the  \(m\)th subcarrier during the  \(n\)th
 symbol interval. The  DD domain symbols are mapped to the TF-domain using the inverse symplectic finite Fourier transform (ISFFT) as described below:
\begin{equation}
\mathbf{X}_{\mathrm{TF}}(m, n) = \frac{1}{\sqrt{MN}}
\sum_{\ell=0}^{M-1}\sum_{c=0}^{N-1}
\mathbf{X}_{\mathrm{DD}}(\ell, c)\, e^{j2\pi\left(\frac{nc}{N}-\frac{m\ell}{M}\right)}.
\label{eq:isfft}
\end{equation}

In matrix form this is compactly written as
\begin{equation}
\mathbf{X}_{\mathrm{TF}} = \mathbf{F}_M\, \mathbf{X}_{\mathrm{DD}}\, \mathbf{F}_N^H,
\label{eq:isfft_matrix}
\end{equation}
where \(\mathbf{F}_M \in \mathbb{C}^{M\times M}\) and \(\mathbf{F}_N \in \mathbb{C}^{N\times N}\) are the unitary discrete Fourier transform (DFT) matrices.

Let \(\mathrm{g}_{\mathrm{tx}}(t)\) be a rectangular transmit pulse of duration \(T\) repeated \(N\) times over the OTFS frame. The Heisenberg transform produces the continuous-time transmit signal represented as 
\begin{equation}
s(t)=\sum_{m=0}^{M-1}\sum_{n=0}^{N-1}
\mathbf{X}_{\mathrm{TF}}(m,n)\,\mathrm{g}_{\mathrm{tx}}(t-nT)\,e^{j2\pi m\Delta f (t-nT)}.
\label{eq:heisenberg}
\end{equation}
Sampling \(s(t)\) at rate \(M/T\) (i.e., with interval \(T/M\)) yields \(MN\) discrete samples given by \(s(\mathrm{g}) = s(t)\big|_{t=\mathrm{g}T/M}\), for \(\mathrm{g} = 0, \dots, MN-1\). The transmitted sample matrix \(\mathbf{S} \in \mathbb{C}^{M \times N}\) may then be expressed as

\begin{equation}
\mathbf{S} = \mathbf{\mathcal{G}}_{\mathrm{tx}}\,\mathbf{F}_M^H \mathbf{X}_{\mathrm{TF}}
= \mathbf{\mathcal{G}}_{\mathrm{tx}}\,\mathbf{X}_{\mathrm{DD}}\,\mathbf{F}_N^H,
\label{eq:transmit_matrix}
\end{equation}
where \(\mathbf{\mathcal{G}}_{\mathrm{tx}} \in \mathbb{C}^{M\times M}\) is a diagonal pulse matrix with \(M\) samples of the transmit pulse. The vectorized transmit block is obtained as \(\mathbf{s}=\mathrm{vec}(\mathbf{S})\in\mathbb{C}^{MN\times 1}\), which can be written as  
\(\mathbf{s} = \big(\mathbf{F}_N^H \otimes \mathbf{\mathcal{G}}_{\mathrm{tx}}\big)\,\mathbf{x}_{\mathrm{DD}}\),  
where \(\mathbf{x}_{\mathrm{DD}}=\mathrm{vec}(\mathbf{X}_{\mathrm{DD}})\in \mathbb{C}^{MN\times 1}\). To mitigate inter-frame interference, a cyclic prefix (CP)
is appended to the transmit vector \(\mathbf{s}\) before transmission.

\subsection{Clustered  DD domain Channel Model}
\label{subsec:channel}
The  DD domain representation of the frequency and time-varying propagation channel is modelled as a superposition of a small number of scatterers and may be written as \cite{hadani2017orthogonal, ramachandran2020otfs}
\begin{equation}
h(\tau,\nu)\;=\;\sum_{i=1}^{L_p} h_i\,\delta(\tau-\tau_i)\,\delta(\nu-\nu_i),
\label{eq:dd_cont}
\end{equation}
where \(L_p\) is the number of dominant multipath components, \(\tau_i\) and \(\nu_i\) are the delay and Doppler shifts of the \(i\)th path, \(h_i\in\mathbb{C}\) denotes its complex path gain and $\delta(\cdot)$ is the Dirac-delta function.
 For typical OTFS parameter choices, the continuous DD shifts associated with the multipath components are mapped onto a discrete DD grid. The delay sampling interval $1/(M\Delta f)$ is typically much smaller than the dominant delay spreads present in wideband systems, which allows path delays to be accurately represented using integer delay taps so that $\tau_i = l_i/(M\Delta f)$ \cite{raviteja2019embedded, srivastava2021bayesian1, raviteja2018interference}. For high-mobility scenarios, however, the Doppler resolution $1/(NT)$ cannot always be assumed to produce Doppler shifts that coincide with discrete grid points. Consequently, the Doppler shift of the $i$th multipath component is modeled as $\nu_i = \frac{c_i}{NT}$, where $ c_i = k_{\nu_i} + \kappa_{\nu_i}$,  $k_{\nu_i} = \mathrm{round}(c_i)$ denotes the nearest integer Doppler tap and $\kappa_{\nu_i}$ represents the associated fractional Doppler component satisfying $|\kappa_{\nu_i}| < \tfrac{1}{2}$.

Although the exact channel distribution may vary across diverse propagation environments, prior studies have shown that clustered multipath propagation is a fundamental characteristic of wideband and high-mobility wireless channels. In such environments, scattering arises from groups of reflectors with distinct physical characteristics, such as spatial location, reflectivity, and mobility, leading to heterogeneous fading behaviours across multipath components. To account for this statistical heterogeneity, GMMs have been widely adopted as a more statistically flexible approximation to practical multipath channels \cite{turan2024wireless,ballal2015low,bock2025sparse}. To the best of our knowledge, in existing OTFS channel modelling, per-path gains are typically assumed to follow a single complex Gaussian distribution for analytical simplicity, which corresponds to the special case of a GMM with  $K = 1$. Therefore, extending this framework to better capture the clustered multipath statistics, the per-path complex gains $h_i$ are modelled as a complex Gaussian-mixture \cite{kong2021variational, selim2015modeling}.
\begin{equation}
p(h_i) = \sum_{k=1}^{K} \rho_k \,\mathcal{CN}\big(\mu_k,\sigma_k^2\big) 
\quad \text{s.t. } \rho_k \ge 0,\quad \sum_{k=1}^K\rho_k=1,
\label{eq:gmm_prior1}
\end{equation}
where \(\mathcal{CN}(\mu_k,\sigma_k^2)\) denotes a circularly-symmetric complex Gaussian with mean \(\mu_k\) and variance \(\sigma_k^2\), while \(K\) is the number of mixture components. Furthermore, $\rho_k$ denotes the mixture weight associated with the  \(k\)th Gaussian component.
Therefore, with a more flexible statistical description of the path gains characterised by these Gaussian–mixture parameters \(\{\rho_k,\mu_k,\sigma_k^2\}\)  in \eqref{eq:gmm_prior1}, the received continuous-time signal is given as 
\begin{equation}
\begin{split}
r(t) &= \iint h(\tau,\nu)\,s(t-\tau)\,e^{j2\pi\nu(t-\tau)}\,d\tau\,d\nu + \eta(t)\\
     &= \sum_{i=1}^{L_p} h_i\,s(t-\tau_i)\,e^{j2\pi\nu_i(t-\tau_i)} + \eta(t),
\end{split}
\label{eq:rcv_cont}
\end{equation}
where \(\eta(t)\) denotes additive white Gaussian noise with mean zero and variance \(\sigma^2\).

Sampling \eqref{eq:rcv_cont} at \(t=\mathrm{g}T/M\) and discarding the initial CP samples yields the discrete-time samples \(r(p)=r(t)\big|_{t=\mathrm{g}T/M}\), \(g=0,\dots,MN-1\), the sampled input–output relationship becomes
\begin{equation}
\begin{split}
r(p) &= \sum_{i=1}^{L_p} h_i\, s\!\big([p-l_i]_{MN}\big)\,
e^{\!\Big(j\frac{2\pi c_i}{MN}(p-l_i)\Big)}\\
&\quad +\; \eta(p), \qquad p=0,\dots,MN-1,
\end{split}
\label{eq:rcv_samples}
\end{equation}
where \([\,\cdot\,]_{MN}\) denotes modulo-\(MN\) indexing (circular shift).
Equation~\eqref{eq:rcv_samples} is the sampled input–output law used for constructing the TD channel matrix $\mathbf{H}$ that maps the transmit sample vector $\mathbf{s} \in\mathbb{C}^{MN\times 1}$ to the receive vector $\mathbf{r}\in\mathbb{C}^{MN\times 1}$ via
\begin{equation}
\mathbf{r} \;=\; \mathbf{H}\,\mathbf{s} + \boldsymbol{\eta},
\label{eq:rs_hw}
\end{equation}
where we have $\mathbf{H} \in\mathbb{C}^{MN\times MN}$ and the noise process obeys $\boldsymbol{\eta} \in\mathbb{C}^{MN\times 1}$. Furthermore, to exploit the DD structure, it is convenient to introduce the standard permutation matrix $\boldsymbol{\Pi}\in\mathbb{C}^{MN\times MN}$ and a diagonal matrix $\boldsymbol{\Delta}\in\mathbb{C}^{MN\times MN}$, where we have
\begin{equation}
\omega = e^{j2\pi c_i/(MN)}, \quad \text{and} \quad
\boldsymbol{\Delta} = \mathrm{diag} \!\big(1,\omega,\omega^{2},\ldots,\omega^{MN-1}\big).
\label{eq:delta_def}
\end{equation}
Note that $\boldsymbol{\Delta}^{c_i}$ imposes a linear phase shift corresponding to Doppler index $c_i$, while $\boldsymbol{\Pi}^{l_i}$ represents the $l_i$-sample forward circular shift \cite{srivastava2021bayesian} upon using these operators, the TD channel matrix is composed as 
\begin{equation}
\mathbf{H} \;=\; \sum_{i=1}^{L_p} h_i\, \boldsymbol{\Pi}^{\,l_i}\,\boldsymbol{\Delta}^{\,c_i}.
\label{eq:H_time}
\end{equation}

The equations \eqref{eq:dd_cont}--\eqref{eq:H_time} preserve the DD parameterisation, i.e. delays, Dopplers, and complex gains, forming the basis for the associated dictionary construction and sparse  DD domain CSI estimation subsequently.

\subsection{OTFS Demodulation}
\label{subsec:demod}

At the receiver, the continuous-time signal is filtered by a matched filter corresponding to \(\mathrm{g}_{\mathrm{rx}}(t)\) and sampled in the TF domain. Let \(\mathbf{R}=\mathrm{vec}^{-1}(\mathbf{r})\in\mathbb{C}^{M\times N}\) denote the received sample matrix. The discrete Wigner transform yields the TF-domain receive matrix
\begin{equation}
\mathbf{Y}_{\mathrm{TF}} = \mathbf{F}_M\mathbf{\mathcal{G}_{rx}}\mathrm{\mathbf{R}},
\label{eq:wigner}
\end{equation}
where \(\mathbf{\mathcal{G}}_{\mathrm{rx}}
= \mathrm{diag}\left\{ \mathrm{g}_{\mathrm{rx}}\!\left(\tfrac{\mathrm{g}T}{M}\right) \right\}_{\mathrm{g}=0}^{M-1}
\). Furthermore, the SFFT maps TF-domain samples to the  DD domain

\begin{equation}
\mathbf{Y}_{\mathrm{DD}}(\ell, c) = \frac{1}{\sqrt{MN}}
\sum_{m=0}^{M-1}\sum_{n=0}^{N-1}
\mathbf{Y}_{\mathrm{TF}}(m, n)\, e^{-j2\pi\left(\frac{nc}{N}-\frac{m\ell}{M}\right)},
\label{eq:fft}
\end{equation}
 where \eqref{eq:fft} can also be written as:
\begin{equation}
\mathbf{Y}_{\mathrm{DD}} = \mathbf{F}_M^H\, \mathbf{Y}_{\mathrm{TF}}\, \mathbf{F}_N = \mathbf{\mathcal{G}}_{\mathrm{rx}}\, \mathbf{R}\, \mathbf{F}_N.
\label{eq:sfft}
\end{equation}
Vectorizing \(\mathbf{Y}_{\mathrm{DD}}\) as \(\mathbf{y}_{\mathrm{DD}}= \mathrm{vec}(\mathbf{Y}_{\mathrm{DD}}) = \big(\mathbf{F}_N \otimes \mathbf{\mathcal{G}}_{\mathrm{rx}}\big)\,\mathbf{r}\), yields the discrete  DD domain linear model
\begin{equation}
\mathbf{y}_{\mathrm{DD}} = \mathbf{H}_{\mathrm{DD}}\,\mathbf{x}_{\mathrm{DD}} + \mathbf{v}_{\mathrm{DD}},
\label{eq:dl_io}
\end{equation}
with
\begin{equation}
\mathbf{H}_{\mathrm{DD}} = (\mathbf{F}_N\otimes \mathbf{\mathcal{G}}_{\mathrm{rx}})\,\mathbf{H}\, (\mathbf{F}_N^H\otimes \mathbf{\mathcal{G}}_{\mathrm{tx}}), \qquad
\mathbf{v}_{\mathrm{DD}} = (\mathbf{F}_N\otimes \mathbf{\mathcal{G}}_{\mathrm{rx}})\,\mathbf{\eta}.
\label{eq:Hdd_vdd}
\end{equation}
The covariance of \(\mathbf{v}_{\mathrm{DD}}\) is formulated as:
\begin{equation}
\mathbf{R}_{v,\mathrm{DD}} = \mathbb{E}\big[\mathbf{v}_{\mathrm{DD}}\mathbf{v}_{\mathrm{DD}}^H\big] = \sigma^2 \big[\mathbf{I}_N\otimes (\mathbf{\mathcal{G}}_{\mathrm{rx}}\mathbf{\mathcal{G}}_{\mathrm{rx}}^H)\big].
\label{eq:rvdd}
\end{equation}
Furthermore, for unit-power transmitted symbols, the linear MMSE detector formulated in the  DD domain is given by
\begin{equation}
\widehat{\mathbf{x}}_{\mathrm{MMSE}}^{\mathrm{DD}}
= \big( \mathbf{H}_{\mathrm{DD}}^{H}\mathbf{R}_{v,\mathrm{DD}}^{-1}\mathbf{H}_{\mathrm{DD}}
+ \mathbf{I}_{MN} \big)^{-1}
\mathbf{H}_{\mathrm{DD}}^{H}\mathbf{R}_{v,\mathrm{DD}}^{-1}\mathbf{y}_{\mathrm{DD}}.
\label{eq:xdd}
\end{equation}
Equations~\eqref{eq:dl_io}--\eqref{eq:xdd} thus establish the discrete  DD domain input–output model of the OTFS system. This compact representation directly exploits the inherent sparsity of the  DD domain channel, providing the foundation for the sparse CSI estimation framework discussed in the next section.

\section{GMM-SBL Based Sparse CSI Estimation for OTFS Systems}
\label{Spare CSI Estimation Model for OTFS Systems}
Let $\mathbf{s}_p \in \mathbb{C}^{N_p \times 1}$ denote the time-domain (TD) pilot vector inserted between OTFS frames for channel estimation. 

The DD channel has a sparse representation 
\begin{equation}
h(\tau,\nu) =
\sum_{i=0}^{M_\tau-1}
\sum_{j=0}^{G_\nu-1}
h_i^j \,
\delta(\tau - \tau_i)\delta(\nu - \nu_j),
\label{eq:dd_sparse}
\end{equation}
where $\tau_i = \frac{i}{M\Delta f}$, and $
\nu_j = \frac{j N_\nu}{G_\nu N T}.$ Here, $M_\tau$ and $N_\nu$ denote the maximum delay and integer-Doppler spreads of the channel satisfying $M_\tau \ll M$ and $N_\nu \ll N$ for a typical under-spread channel. As discussed before, since the delay resolution is sufficiently high, it is adequate to represent delays using integer delay taps. Whereas, to accurately capture fractional Doppler shifts, the Doppler grid size is chosen as $G_\nu \gg N_\nu$, thereby refining the Doppler sampling resolution. In this formulation, the integer Doppler tap $k_{\nu_j} = \mathrm{round}\left( \frac{jN_{\nu}}{G_{\nu}} \right)$, and the fractional Doppler is represented by $\kappa_{\nu_j} =
\frac{jN_{\nu}}{G_{\nu}} - \mathrm{round}\left( \frac{jN_{\nu}}{G_{\nu}} \right)$.
Since only a few dominant reflectors exist, the DD channel remains sparse over the $M_\tau G_\nu$ grid, with $L_p \ll M_\tau G_\nu$ non-zero coefficients.\\
After CP removal, the received TD pilot vector is expressed as
\begin{equation}
\mathbf{r}_p = \mathbf{H}\mathbf{s}_p + \boldsymbol{\eta}_p,
\label{eq:pilot_io}
\end{equation}
where $\boldsymbol{\eta}_p \in \mathbb{C}^{N_p \times 1}$ denotes additive white Gaussian noise.

The effective channel $\overline{\mathbf{H}}$ can be written as
\begin{equation}
\overline{\mathbf{H}} = \sum_{i=0}^{M_\tau-1}\sum_{j=0}^{G_\nu-1} h_i^j\, (\overline{\boldsymbol{\Pi}})^i\, (\overline{\boldsymbol{\Delta}}_i)^j,
\end{equation}
with $\overline{\boldsymbol{\Pi}} \in \mathbb{C}^{N_p \times N_p}$ a permutation matrix 
and $\overline{\boldsymbol{\Delta}}_i \in \mathbb{C}^{N_p \times N_p}$ a diagonal matrix given by

\begin{equation}
\vcenter{\hbox{$
\overline{\boldsymbol{\Delta}}_i =
\begin{cases}
\operatorname{diag}(\omega^0, \omega^1, \ldots, \omega^{N_p-i-1}, \omega^{-i}, \ldots, \omega^{-1}), & \text{if } i \neq 0,\\
\operatorname{diag}(\omega^0, \omega^1, \ldots, \omega^{N_p-1}), & \text{if } i = 0,
\end{cases}
$}}
\label{eq:delta_i}
\end{equation}
where $\omega = e^{j2\pi\frac{N_{\nu}}{G_{\nu}MN}}$. Each $\boldsymbol{\omega}_{i}^{j}$ represents the contribution of the pilot sequence $\mathbf{s}_p$ as it is 
shifted in delay by $i$ samples (via $(\overline{\boldsymbol{\Pi}})^i$) and modulated in Doppler by $j$ shifts 
(via $(\overline{\boldsymbol{\Delta}}_i)^j$). 
Thus, $\boldsymbol{\omega}_{i}^{j}$ can be interpreted as the effective measurement vector corresponding to the 
 \((i,j)\)th DD bin. 
By stacking all such vectors column-wise, we obtain the dictionary matrix $\boldsymbol{\Omega}$ and the sparse vector $\mathbf{h}$  as
\begin{align}
\boldsymbol{\Omega} &= \left[ \boldsymbol{\omega}^0_0, \boldsymbol{\omega}^1_0, \ldots, \boldsymbol{\omega}^{N_\nu-1}_0, \boldsymbol{\omega}^0_1, \ldots, \boldsymbol{\omega}^{G_\nu-1}_{M_\tau-1} \right] \\
\mathbf{h} &= \left[ h^0_0, h^1_0, \ldots, h^{N_\nu-1}_0, h^0_1, \ldots, h^{G_\nu-1}_{M_\tau-1} \right]^T.
\end{align}
The pilot observation model then becomes
\begin{equation}
\mathbf{r}_p = \mathbf{\Omega}\mathbf{h} + \boldsymbol{\eta}_p,
\label{eq:sparse_model}
\end{equation}
which forms the basis for the subsequent GMM-SBL-based sparse  DD domain channel estimation.

The  DD domain channel response in OTFS systems is characterised by a sparse structure arising from a finite number of dominant multipath components. However, in practical scenarios, the complex gains associated with these components may exhibit non-stationary statistics, often organised into distinct clusters due to the presence of grouped scattering objects. This gives rise to a multimodal distribution of tap gains, a feature that is not captured by conventional estimators relying on a homogeneous single Gaussian prior.

To better model this statistical heterogeneity and inherent sparsity in the  DD domain, a GMM-SBL framework
is proposed for sparse CSI estimation. The proposed framework introduces a more expressive, hierarchical prior that models the channel as a mixture of $K$ Gaussian components
 \begin{equation}
p(\mathbf{h}) = \sum_{k=1}^{K} \rho_k \, \mathcal{CN}(\boldsymbol{\mu}_k,\boldsymbol{\Gamma}_k),
\qquad 
\sum_{k=1}^{K} \rho_k = 1,\;\rho_k \geq 0,
\label{eq:gmmprior}
\end{equation}
where $\boldsymbol{\Gamma}_k = \text{diag}(\boldsymbol{\gamma}_k)$ is the diagonal covariance matrix for the \(k\)th component of the channel $\mathbf{h}$. Also, the corresponding marginal covariance matrix is $\mathbf{A}_k$. The posterior mean and covariance under component $k$ are denoted by $\boldsymbol{\mu}_k$ and $\boldsymbol{\Sigma}_k$, respectively. The responsibility is written as $\pi_k(\mathbf{r}_{P})=\Pr(k\mid\mathbf{r}_{P})$, and $\pi_{i,k}$ in the multi-snapshot setting.

The GMM--SBL approach represents a significant paradigm shift from imposing a sparse structure to enabling data-driven learning, as it allows the model to capture diverse channel statistics within a unified Bayesian framework. A critical advantage of this formulation is its strong theoretical foundation \cite{bock2024sparse}, which exhibits sparsity-inducing properties, as shown in Theorem \ref{gmm-sbl-sparse}.

\begin{theorem}[\textbf{Sparsity of the GMM-SBL Prior}]
 \label{gmm-sbl-sparse}
Let the latent variable $z \in \{1, \ldots, K\}$ indicate the mixture 
component with probability $\Pr(z{=}k) = \rho_k$. 
Consider the hierarchical prior
\[
p(\mathbf{h}) \;=\; \sum_{k=1}^{K} \rho_k \, 
\mathcal{CN}\!\big( \mathbf{0}, \boldsymbol{\Gamma}_k\big),
\]
where $\boldsymbol{\Gamma}_k = \mathrm{diag}(\boldsymbol{\gamma}_k)$ 
with $\boldsymbol{\gamma}_k > 0$. 
Then there exists a constant $C > 0$ so that
\begin{equation}
p(\mathbf{h}) \;\leq\; C \cdot \prod_{r=1}^{M_\tau N_\nu} \frac{1}{|\;h_r\;|^{2}}.
\end{equation}
\noindent\textit{Proof:} Given in Appendix \ref{appendix a}.
\end{theorem}

This result ensures that the GMM-SBL prior is bounded by a sparsity-promoting function \cite{wipf2004sparse}. For our problem, let the function be $t(h)$ obey $t(h) \propto \prod_r |h(r)|^{-2}$. Consequently, evidence maximisation within the proposed framework is guaranteed to promote the robust recovery of sparse solutions, while providing the flexibility to learn the parameters ${\boldsymbol{\gamma}_k, \rho_k}$ and adapt to the underlying channel structure.

\noindent The GMM-SBL framework models the  DD domain channel vector
with a hierarchical prior that adaptively learns sparsity patterns from data. Given the received pilots, the posterior distribution conditioned on mixture component \(k\)
is Gaussian, and it is defined as 

\begin{equation}
p(\mathbf{h}\mid \mathbf{r}_p,k)=\mathcal{CN}\!\big(\boldsymbol{\mu}_k,\boldsymbol{\Sigma}_k\big),
\end{equation}
with the posterior mean $\boldsymbol{\mu}_k$ and posterior covariance $\boldsymbol{\Sigma}_k$ under component $k$ defined as
\begin{align}
\mathbf{A}_k &= \sigma^2 \mathbf{I} + \boldsymbol{\Omega}\,\boldsymbol{\Gamma}_k\,\boldsymbol{\Omega}^H, \label{eq:Ak}\\
\boldsymbol{\mu}_{k} &= \boldsymbol{\Gamma}_k\,\boldsymbol{\Omega}^H \mathbf{A}_k^{-1}\mathbf{r}_p, \label{eq:muk}\\
\boldsymbol{\Sigma}_k &= \boldsymbol{\Gamma}_k - \boldsymbol{\Gamma}_k\,\boldsymbol{\Omega}^H\mathbf{A}_k^{-1}\boldsymbol{\Omega}\,\boldsymbol{\Gamma}_k. \label{eq:Sigmak}
\end{align}
The marginal likelihood under component $k$ is 
\[
p(\mathbf{r}_p \mid k)=\mathcal{CN}(\mathbf{0},\mathbf{A}_k).
\]
The responsibility $\pi_k(\mathbf{r}_P)$ is the posterior probability that the $k$th mixture component generated the received pilot vector $\mathbf{r}_p$; according to the Bayes' rule
\[
\pi_k(\mathbf{r}_P)=\Pr\big(k\mid\mathbf{r}_P\big)
=\frac{\rho_k\,p(\mathbf{r}_P\mid k)}{\sum_{\ell=1}^K \rho_\ell\,p(\mathbf{r}_P\mid \ell)},
\]
so $0\le \pi_k(\mathbf{r}_P)\le 1$ and $\sum_{k=1}^K \pi_k(\mathbf{r}_P)=1$. The responsibilities provide a probabilistic assignment of observations to components and quantify each component’s contribution to the posterior distribution \cite{deisenroth2020mathematics}. As will be seen in subsequent parameter updates, they act as adaptive weights for learning $\boldsymbol{\Gamma}_k$ and $\rho_k$ from the aggregated posterior means and variances. 

Also, the conditional-mean estimate (CME) is given by
\[
\widehat{\mathbf{h}}=\sum_{k=1}^K \pi_k(\mathbf{r}_p)\,\boldsymbol{\mu}_k.
\]

\begin{theorem}[\textbf{EM updates for GMM--SBL}]
\label{th:EM_final}
Given $L$ pilot observations $\{\mathbf{r}_{p,i}\}_{i=1}^L$, the EM updates at iteration $t$ are given by

\medskip
\noindent\textbf{E-step.} For each snapshot $i$ and mixture component $k$ compute the posterior mean
\[
\label{eq:mu_em_final}
\boldsymbol{\mu}_{i,k}^{(t)} \;=\; \boldsymbol{\Gamma}_{k}^{(t)}\boldsymbol{\Omega}^H\big(\mathbf{A}_{k}^{(t)}\big)^{-1}\,\mathbf{r}_{p,i},
\]

and the posterior covariance
\[
\boldsymbol{\Sigma}_{k}^{(t)} \;=\; \boldsymbol{\Gamma}_{k}^{(t)} 
\;-\; \boldsymbol{\Gamma}_{k}^{(t)}\boldsymbol{\Omega}^H\big(\mathbf{A}_{k}^{(t)}\big)^{-1}\boldsymbol{\Omega}\,\boldsymbol{\Gamma}_{k}^{(t)},
\]
where we have
\[
\label{eq:Ak_em_final}
\mathbf{A}_{k}^{(t)} \;=\; \sigma^2\mathbf{I} \;+\; \boldsymbol{\Omega}\,\boldsymbol{\Gamma}_{k}^{(t)}\,\boldsymbol{\Omega}^H.
\]
Form the responsibility
\[
\pi_{i,k}^{(t)} \;=\; 
\frac{\rho_k^{(t)}\,p\big(\mathbf{r}_{p,i}\mid k;\mathbf{A}_k^{(t)}\big)}
{\sum_{\ell=1}^K \rho_\ell^{(t)}\,p\big(\mathbf{r}_{p,i}\mid \ell;\mathbf{A}_\ell^{(t)}\big)},
\]
\textit{where we have}
\[
p\big(\mathbf{r}_{p,i}\mid k;\mathbf{A}_k^{(t)}\big)=\mathcal{CN}(\mathbf{0},\mathbf{A}_k^{(t)}),
\qquad
\mathbf{A}_k^{(t)}=\sigma^2 \mathbf{I} + \boldsymbol{\Omega}\,\boldsymbol{\Gamma}_k^{(t)}\,\boldsymbol{\Omega}^H.
\]
\medskip
\textbf{M-step.} For $r=1,\ldots,M_\tau N_\nu$ update the hyperparameters $\gamma_{k,r}$ and the mixture weights $\rho_k$, where we have:
\begin{align}
\gamma_{k,r}^{(t+1)}
&=
\frac{\displaystyle\sum_{i=1}^L \pi_{i,k}^{(t)}\Big(|\mu_{i,k,r}^{(t)}|^2+\Sigma_{k,r,r}^{(t)}\Big)}
{\displaystyle\sum_{i=1}^L \pi_{i,k}^{(t)}}, \label{eq:gamma_update_final}\\[6pt]
\rho_k^{(t+1)}
&=
\frac{1}{L}\sum_{i=1}^L \pi_{i,k}^{(t)}. \label{eq:rho_update_final}
\end{align}
\noindent\textit{Proof:} Given in Appendix \ref{appendix b}.
\end{theorem}
After convergence (at iteration $T$), the CME for each snapshot is obtained as:
\[
\widehat{\mathbf{h}}^{(i)} \;=\; \sum_{k=1}^K \pi_{i,k}^{(T)}\,\boldsymbol{\mu}_{i,k}^{(T)}, 
\qquad i=1,\dots,L.
\]

Although the proposed formulation employs a Gaussian mixture prior, sparsity is enforced through the relevance determination mechanism inherent to sparse Bayesian learning. In conventional SBL, a single Gaussian prior is combined with hyperparameter-driven variance adaptation to promote sparse solutions. By contrast, mixture-prior approaches typically incorporate mixture structures directly at the coefficient level. The proposed GMM-SBL framework preserves the explicit hyperparameter based sparsity control of SBL while introducing mixture modeling to capture statistical heterogeneity across channel realizations. The posterior responsibilities act as adaptive weights in the hyperparameter updates and do not impose coefficientwise hard clustering of DD taps. Consequently, the method extends the modeling capability of SBL, while retaining its underlying EM-based learning structure.
Algorithm-\ref{algo1} details the steps of GMM-SBL based channel estimation for the OTFS system considered.

\begin{algorithm}[t]
\scriptsize
\caption{GMM--SBL Based CSI Estimation in OTFS}
\label{algo1}

\textbf{Input:} Snapshots $\mathbf{R}' = \{ \mathbf{r}_{p,i} \}_{i=1}^L$, dictionary matrix $\boldsymbol{\Omega} \in \mathbb{C}^{N_p \times M_\tau N_\nu}$, noise variance $\sigma^2$, mixture components $K$, EM iterations $T$. \\[3pt]
\textbf{Output:} $\{\rho_k^{(T)}\}_{k=1}^K$, $\{\boldsymbol{\Gamma}_k^{(T)}\}_{k=1}^K$, $\{\hat{\mathbf{h}}_i\}_{i=1}^L$. \\[3pt]

\textbf{Initialize:} $\rho_k^{(0)} = 1/K$, $\boldsymbol{\Gamma}_k^{(0)} = \mathbf{I}_{M_\tau N_\nu}$.

\BlankLine
\SetAlgoNlRelativeSize{-1}

\For{$t \leftarrow 0$ \KwTo $T-1$}{
    \For{$k \leftarrow 1$ \KwTo $K$}{
        $\mathbf{A}_k^{(t)} = \sigma^2 \mathbf{I}_{N_p} + \boldsymbol{\Omega} \boldsymbol{\Gamma}_k^{(t)} \boldsymbol{\Omega}^H$. \\
        \For{$i \leftarrow 1$ \KwTo $L$}{
            $\mathbf{u}_{i,k}^{(t)} = \mathbf{A}_k^{(t)^{-1}} \mathbf{r}_{p,i}$. \\
            $\boldsymbol{\mu}_{i,k}^{(t)} = \boldsymbol{\Gamma}_k^{(t)} \boldsymbol{\Omega}^H \mathbf{u}_{i,k}^{(t)}$. \\
            $\boldsymbol{\Sigma}_{k}^{(t)} = \boldsymbol{\Gamma}_k^{(t)} - \boldsymbol{\Gamma}_k^{(t)} \boldsymbol{\Omega}^H \mathbf{A}_k^{(t)^{-1}} \boldsymbol{\Omega} \boldsymbol{\Gamma}_k^{(t)}$. \\
            $\ell_{i,k}^{(t)} = -\big(\mathbf{r}_{p,i}^H \mathbf{A}_k^{(t)^{-1}} \mathbf{r}_{p,i} + \log\det \mathbf{A}_k^{(t)} + N_p \log(\pi)\big)$. \\
        }
    }
    \For{$i \leftarrow 1$ \KwTo $L$}{
        $\tilde{\pi}_{i,k}^{(t)} = \rho_k^{(t)} e^{\ell_{i,k}^{(t)}}, \quad k = 1,\dots,K$. \\
        $\pi_{i,k}^{(t)} = \tilde{\pi}_{i,k}^{(t)} \Big/ \sum_{r=1}^K \tilde{\pi}_{i,r}^{(t)}$. \\
    }
    \For{$k \leftarrow 1$ \KwTo $K$}{
        $N_k^{(t)} = \sum_{i=1}^L \pi_{i,k}^{(t)}$. \\
        $\mathbf{e}_{i,k}^{(t)} = |\boldsymbol{\mu}_{i,k}^{(t)}|^2 + \operatorname{diag}(\boldsymbol{\Sigma}_{k}^{(t)})$. \\
        $\boldsymbol{\Gamma}_k^{(t+1)} = \operatorname{diag}\!\left(\frac{1}{N_k^{(t)}} \sum_{i=1}^L \pi_{i,k}^{(t)} \mathbf{e}_{i,k}^{(t)} \right)$. \\
        $\rho_k^{(t+1)} = N_k^{(t)} / L$. \\
    }
}

\For{$i \leftarrow 1$ \KwTo $L$}{
    $\hat{\mathbf{h}}_i = \sum_{k=1}^K \pi_{i,k}^{(T)} \boldsymbol{\mu}_{i,k}^{(T)}$. \\
}

\end{algorithm}

\section{Performance Benchmarks}
\label{Performance Benchmark}

To assess the estimator's performance, a pair of reference bounds is employed. 
First is an Oracle–MMSE (genie estimator) assumes perfect knowledge of 
the  DD domain support set $\mathcal{H}$. 
Let $\boldsymbol{\Omega}_O=\boldsymbol{\Omega}(:,\mathcal{H})$ denote the Oracle sensing matrix.
The Oracle–MMSE estimate is formulated as:
\[
\widehat{\mathbf{h}}_{\mathrm{O\text{-}MMSE}}
\;=\;
\big(\boldsymbol{\Omega}_O^{H}\mathbf{R}_v^{-1}\boldsymbol{\Omega}_O + \mathbf{I}\big)^{-1}
\boldsymbol{\Omega}_O^{H}\mathbf{R}_v^{-1}\,\mathbf{r}_{p} .
\]

The simulated Oracle implements the analytic bound by applying MMSE on the true support to
$\mathbf{r}_{p}$ and averaging the normalised squared error over Monte Carlo trials.
The second benchmark considered is the BCRLB, which captures fundamental limits when a prior $p(\mathbf{h})$ is available \cite{trees2007bayesian, srivastava2021bayesian}.
Given $L$ snapshots,
the Bayesian information decomposes as $\mathbf{J}=\mathbf{J}_{\mathrm{data}}+\mathbf{J}_{\mathrm{prior}}$ with
$\mathbf{J}_{\mathrm{data}}=L\,\boldsymbol{\Omega}^{H}\mathbf{R}_v^{-1}\boldsymbol{\Omega}$ and
$\mathbf{J}_{\mathrm{prior}}=-\mathbb{E}_{\mathbf{h}}\!\big[\nabla_{\mathbf{h}}^2\log p(\mathbf{h})\big]$.
For a general $K$-component complex Gaussian mixture prior, $\mathbf{J}_{\mathrm{prior}}$ has no closed form \cite{crafts2024bayesian}. 
The BCRLB for a single-Gaussian prior $p(\mathbf{h})=\mathcal{CN}(\mathbf{0},\boldsymbol{\Gamma})$ can be expressed as (see Appendix~\ref{appendix c} for derivation)
\begin{equation}
\mathrm{MSE}(\hat{\mathbf{h}}) \;\ge\;
\mathrm{tr}\!\left( \big[ \mathbf{J}_{\mathrm{data}} + \mathbf{J}_{\mathrm{prior}} \big]^{-1} \right).
\label{eq:bcrlb_part1}
\end{equation}

The prior term reduces to $\mathbf{J}_{\mathrm{prior}}=\boldsymbol{\Gamma}^{-1}$, and the BCRLB admits the closed form shown above 
when the mixture components share uniform variance. Moreover, when $L$ is large, the data term dominates, and the
BCRLB coincides with the Oracle reference; otherwise the Monte Carlo–evaluated $\mathbf{J}_{\mathrm{prior}}$ captures the effect of the GMM prior on the fundamental MSE limit.

\begin{table}[!t]
\footnotesize
\renewcommand{\arraystretch}{1.1} 
\caption{Simulation Parameters}
\centering
\label{sub6parameters}
\begin{tabular}{|l|l|}
\hline
\multicolumn{1}{|c|}{\textbf{Parameter (Symbol)}} & \multicolumn{1}{c|}{\textbf{Value}} \\ \hline
Carrier Frequency in GHz ($f_c$) & $4$ \\ \hline
Subcarrier spacing in kHz ($\Delta f$) & $15$ \\ \hline
No. of symbols along delay-axis ($M$) & $32$ \\ \hline
No. of symbols along Doppler-axis ($N$) & $32$ \\ \hline
Max. spread across delay-axis ($M_\tau$) & $16$ \\ \hline
Max. spread across Doppler-axis ($N_\nu$) & $10$ \\ \hline
Doppler grid size ($G_\nu$) & $20$ \\ \hline
No. of samples in CP ($P$) & $16$ \\ \hline
No. of pilots ($N_p$) & $80$ \\ \hline
No. of dominant reflectors ($L_p$) & $5$ \\ \hline
Modulation scheme & QPSK \\ \hline
No. of training snapshots ($L$) & $10$ \\ \hline
Pulse-shape & Rectangular \\ \hline
\end{tabular}
\end{table}

\begin{table}[!t]
\footnotesize
\renewcommand{\arraystretch}{1.1} 
\caption{DD-Profile of the Wireless Channel}
\centering
\label{ddprofile}
\begin{tabular}{|l|c|c|c|c|c|}
\hline
Path-index ($i$) & 1 & 2 & 3 & 4 & 5 \\ \hline
Delay in $\mu$s ($\tau_i$) & 2.08 & 4.164 & 6.246 & 8.328 & 10.42 \\ \hline
Doppler in Hz ($\nu_i$) & 0 & 470 & 940 & 1410 & 1880 \\ \hline
Speed in km/h & 0 & 126.9 & 253.8 & 380.7 & 507.6 \\ \hline
\end{tabular}
\end{table}

\section{Results and Discussion}
\label{Results and Discussion}

This section presents a comprehensive performance evaluation of the proposed GMM–SBL scheme for sparse  DD domain channel estimation. The normalised mean square error (NMSE) and symbol error rate (SER) are used as  performance metrics, where the NMSE is defined as
\begin{equation}
\text{NMSE} = \frac{\lVert \hat{\mathbf{H}}_{\mathrm{DD}} - \mathbf{H}_{\mathrm{DD}} \rVert^2}
{\lVert \mathbf{H}_{\mathrm{DD}} \rVert^2},
\end{equation}
where $\hat{\mathbf{H}}_{\text{DD}}$ and $\mathbf{H}_{\text{DD}}$ denote the estimated and the true  DD domain channel matrices, respectively, calculated from (\ref{eq:Hdd_vdd}).

\subsection{Simulation Parameters}
The OTFS system parameters (including $M$, $N$, $M_{\tau}$, and $N_{\nu}$) are selected to reflect the high-mobility scenarios and to satisfy the underspread channel assumption commonly adopted in OTFS literature \cite{hong2022delay, srivastava2021bayesian, srivastava2021bayesian1, mohammed2024otfs}. The grid sizes $M \in \{32, 64\}$ chosen are appropriate for the considered bandwidth and delay spread, enabling accurate representation of multipath delays on the integer delay grid. Similarly, $N = 32$, together with the fractional grid formulation, enables accurate modeling of fractional Doppler effects. The delay and Doppler supports $M_{\tau}$ and $N_{\nu}$ are selected to capture the effective channel spread while maintaining $M_{\tau} \ll M$ and $N_{\nu} \ll N$, consistent with practical OTFS system design.\\
To ensure fairness, the delay and Doppler indices are randomly generated for each Monte Carlo trial, and the pilot power is normalised across all schemes. The maximum number of EM iterations is set to $100$ for both GMM–SBL and SBL. For  Least absolute shrinkage and selection operator (LASSO) a fixed sparsity regularisation parameter of $10^{-3}$ is used. Focal underdetermined system solver (FOCUSS)  employs an $\ell_p$ norm with $p=0.8$, a noise variance–based regularization, a stopping tolerance of $10^{-6}$, and a maximum of 500 iterations. The OMP algorithm uses a residual-based stopping rule, terminating once the change in residual error falls below a noise-dependent threshold, which we set to $10^{-2}$.

These parameter values were selected after preliminary evaluation across the considered SNR range to identify stable operating settings for each algorithm. In particular, the regularisation parameter for LASSO, the $\ell_p$-norm parameter $p$ and stopping tolerance for FOCUSS, as well as the residual-based threshold for OMP were identified to ensure reliable convergence behaviour and consistent estimation performance under the specified channel conditions. For SBL and GMM--SBL, the maximum number of EM iterations was set identically to provide comparable convergence conditions. For each benchmark method, the results reported correspond to the configuration that demonstrated stable performance across Monte Carlo trials, thereby enabling a meaningful comparison with the proposed approach.

\begin{figure*}[!t]
  \centering
  \begin{subfigure}[b]{0.48\linewidth}
    \centering
    \includegraphics[height=7.5cm]{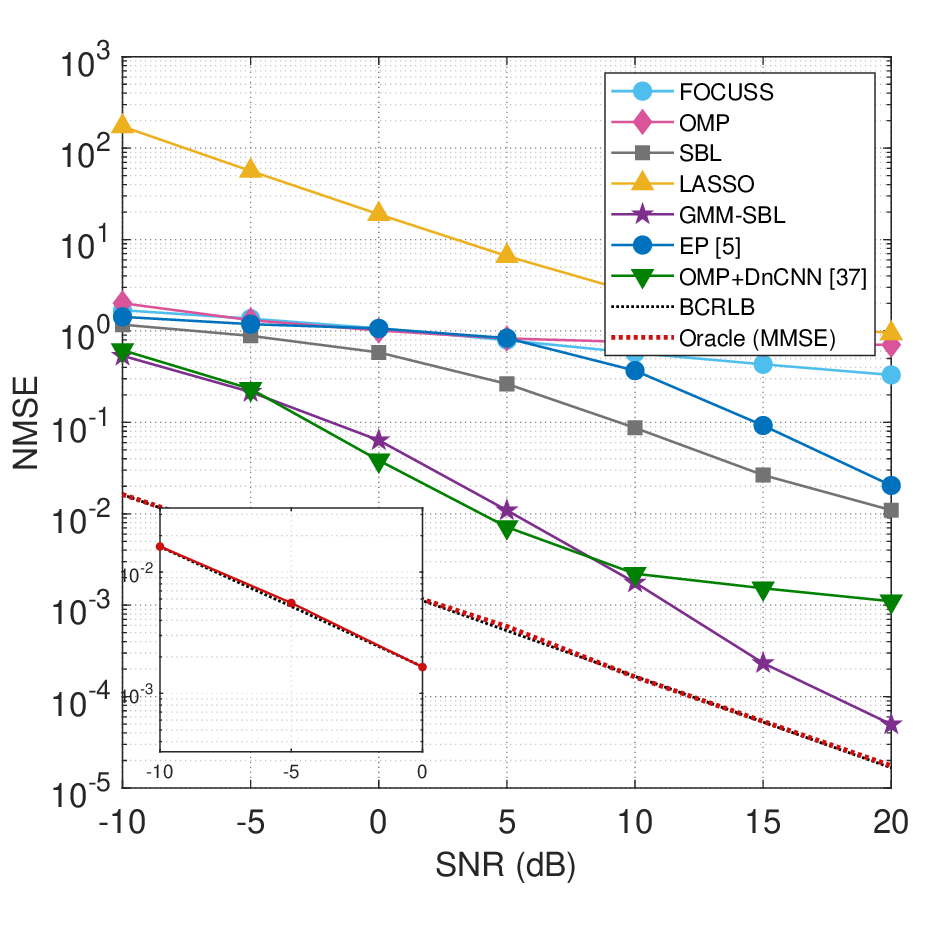}
    \caption{ NMSE vs SNR.}
    \label{fig:comp}
  \end{subfigure}\hfill
  \begin{subfigure}[b]{0.46\linewidth}
    \centering
    \includegraphics[height=7.5cm]{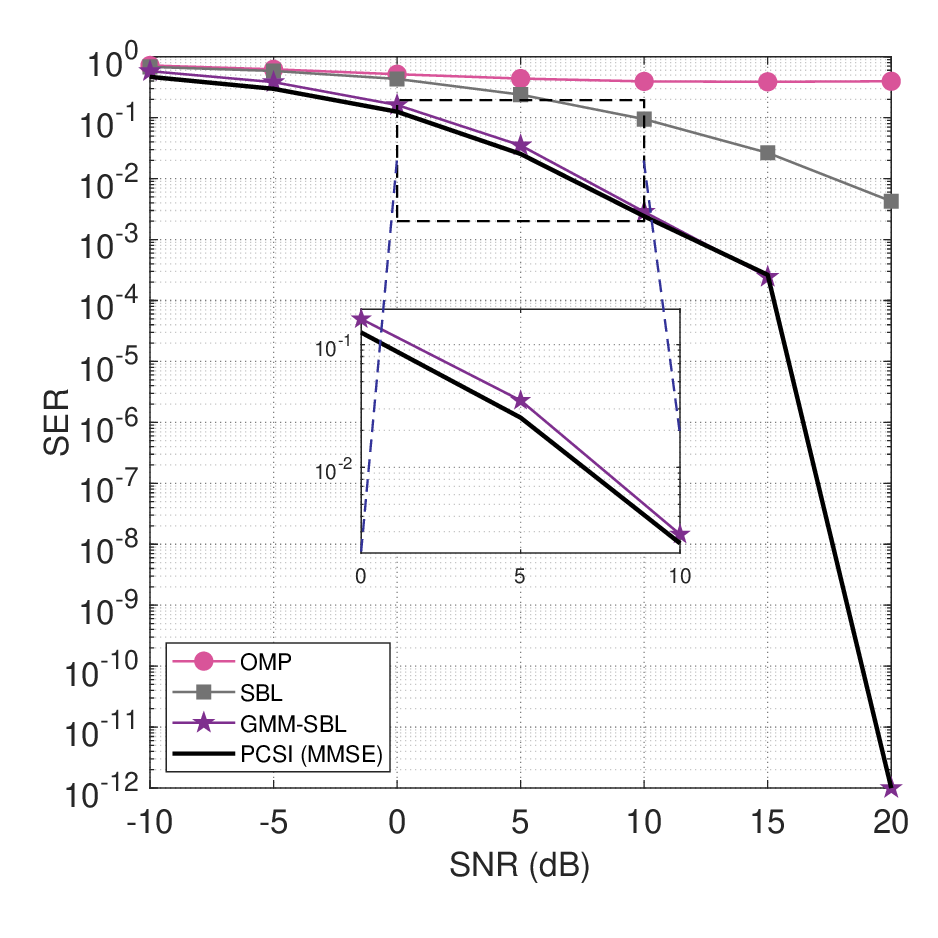}
    \caption{SER vs SNR.}
    \label{fig:ser}
  \end{subfigure}

  \vspace{2mm}
  \caption{Performance comparison of proposed GMM-SBL algorithm with benchmarks and existing state-of-the-art methods.}
  \label{fig:comparison}
\end{figure*}

\subsection{Performance Comparison}
The method is compared to state-of-the-art compressed sensing techniques, including OMP \cite{cai2011orthogonal}, FOCUSS \cite{gorodnitsky2002sparse}, SBL \cite{wipf2004sparse}, and LASSO \cite{ranstam2018lasso}. In addition, performance is evaluated against the embedded pilot [EP] based OTFS channel estimation scheme \cite{raviteja2019embedded}, which serves as a classical benchmark, as well as a recent DL-based approach that combines OMP with a denoiser \cite{he2023denoising}. As observed in Fig. \ref{fig:comp}, the EP based OTFS scheme achieves reliable performance; however, it relies on DD domain impulse pilots, guard regions, and threshold based detection. By contrast, the improved GMM--SBL framework conceived employs TD pilots and performs Bayesian inference under a learned statistical prior, thereby avoiding threshold tuning, while adaptively exploiting clustered DD sparsity.\\

The DL-based method achieves competitive performance in the low-SNR regime, but at the cost of substantial training overhead, which is not possible in high mobility scenarios. Nevertheless, its performance saturates at higher SNR values, as the denoiser is primarily trained for noise suppression rather than for modelling the underlying sparse channel structure, limiting further improvement when the noise becomes negligible. Classical sparse recovery methods exhibit additional limitations. OMP is sensitive to the stopping criterion, FOCUSS may suffer from slow or unstable convergence, and LASSO experiences estimation bias under DD domain sparsity. Conventional SBL improves robustness through hyperparameter learning; however, its single-Gaussian prior restricts statistical flexibility. By incorporating mixture-level statistical modelling within the SBL framework, the proposed GMM--SBL achieves consistently lower NMSE across the evaluated SNR range.\\ 
\begin{figure*}[!h]
  \centering
  \begin{subfigure}[b]{0.48\linewidth}
    \centering
    \includegraphics[height=7.5cm]{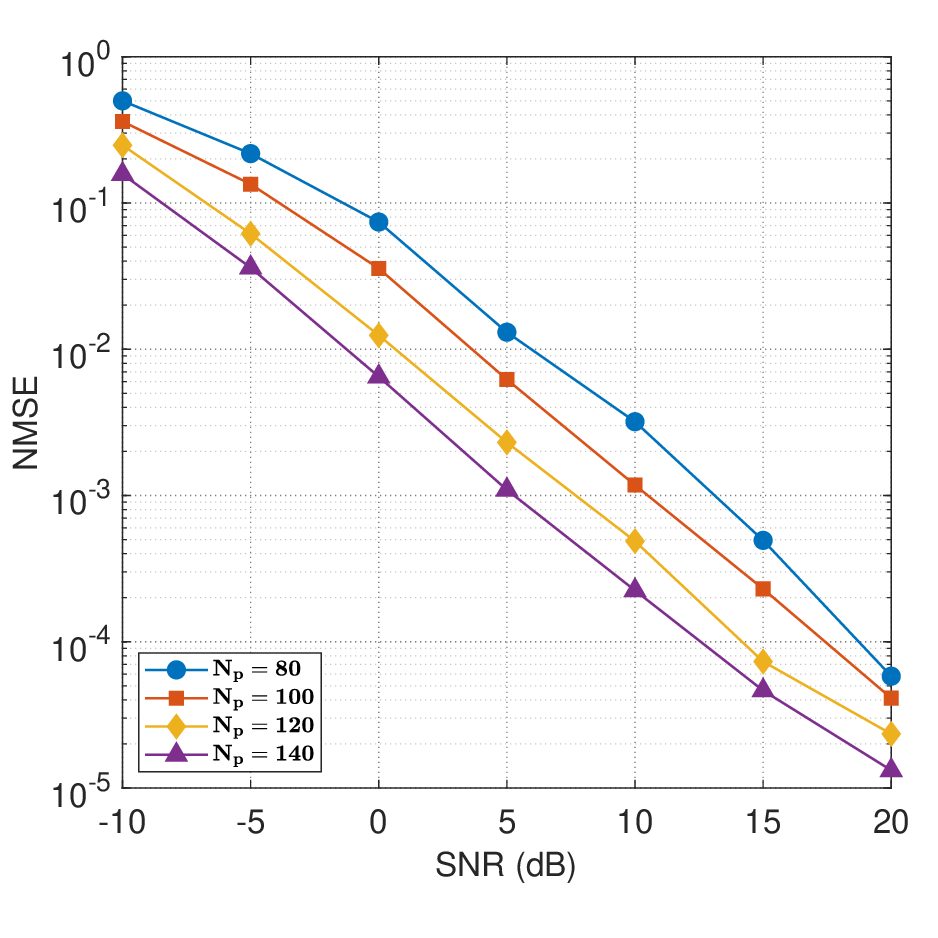}
    \caption{NMSE vs SNR for different pilot overhead.}
    \label{fig:np}
  \end{subfigure}\hfill
  \begin{subfigure}[b]{0.48\linewidth}
    \centering
    \includegraphics[height=7.5cm]{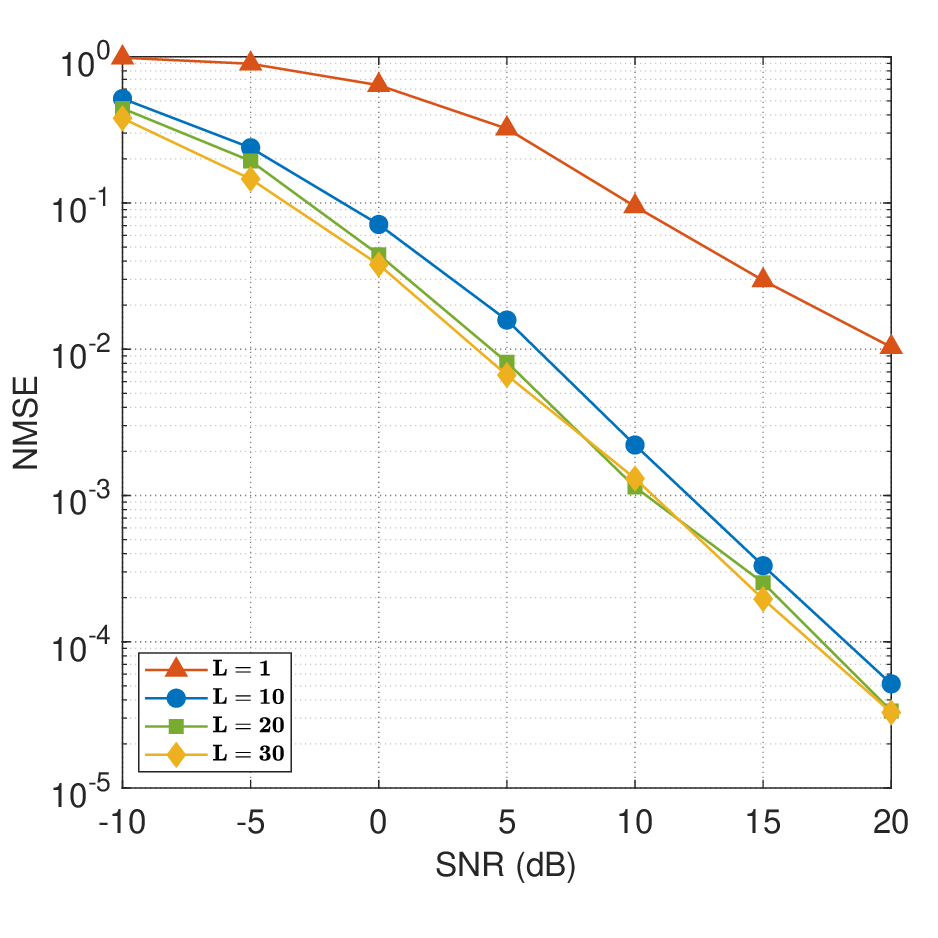}
    \caption{NMSE vs SNR for different No. of snapshots $L$.}
    \label{fig:L}
  \end{subfigure}

  \vspace{2mm}
  \caption{Analysis of proposed GMM-SBL algorithm with different parameters.}
  \label{fig:np_L}
\end{figure*}
Furthermore, the proposed estimator is benchmarked against the Oracle-MMSE and the BCRLB derived in Section \ref{Performance Benchmark}. As shown in Fig. \ref{fig:comp} and Fig. \ref{fig:ser}, GMM–SBL consistently outperforms all baselines, and at high SNR its NMSE approaches the Oracle–MMSE and BCRLB, demonstrating near-optimal estimation accuracy. These NMSE improvements directly translate into SER gains, where GMM–SBL achieves near-perfect CSI detection while OMP and FOCUSS exhibit significant performance degradation

\begin{figure*}[!t]
  \centering
  \begin{subfigure}[b]{0.48\linewidth}
    \centering
    \includegraphics[height=7.5cm]{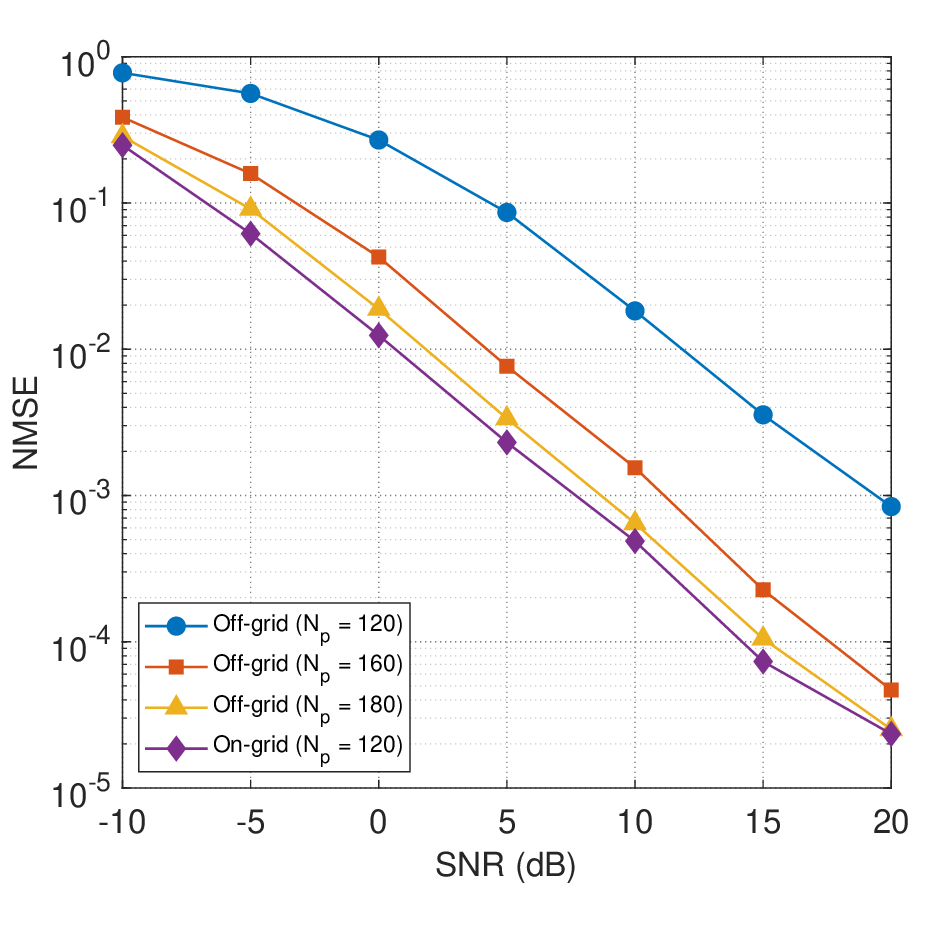}
    \caption{ NMSE vs SNR with fractional Doppler conditions.}
    \label{fig:frac_Doppler}
  \end{subfigure}\hfill
  \begin{subfigure}[b]{0.46\linewidth}
    \centering
    \includegraphics[height=7.5cm]{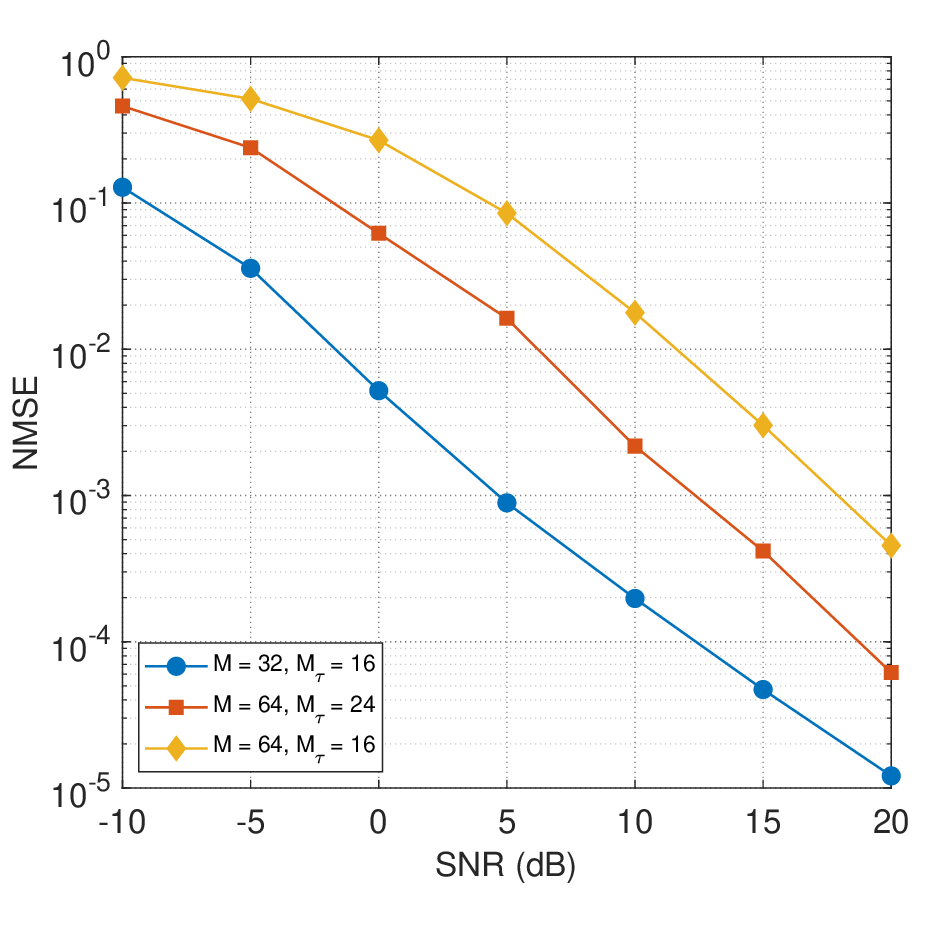}
    \caption{NMSE vs SNR for different delay resolutions.}
    \label{fig:scaled_M}
  \end{subfigure}

  \vspace{2mm}
  \caption{Impact of DD grid resolution on GMM-SBL channel estimation performance.}
  \label{fig:dd_grid}
\end{figure*}

\subsection{Parameter Sensitivity Analysis}
The robustness of GMM–SBL is further evaluated under varying pilot length and number of snapshots. Fig. \ref{fig:np} shows that increasing the number of pilot symbols $N_p$ significantly reduces the NMSE. We observe more than $6\ dB$ performance improvement upon increasing $N_p$ from $80$ to $140$. Morever, it can be seen in Fig. \ref{fig:L} that additional snapshots provide improved NMSE performance. Explicitly, at $0$ dB the NMSE decreases from $6.39 \times 10^{-1}$ with $L=1$ to $7.13 \times 10^{-2}$ with $L=10$.

Furthermore, to evaluate the sensitivity of the proposed method under realistic mobility conditions, 
fractional Doppler shifts are incorporated in the channel model. Specifically, the normalized Doppler index of each multipath component is includes an integer Doppler tap $k_{\nu_i}$, while the fractional Doppler term $\kappa_{\nu_i}$ is uniformly generated in the range $\left(-\tfrac{1}{2}, \tfrac{1}{2}\right)$. To accurately capture these off-grid components, the grid resolution was increased such that $G_\nu \gg N_\nu$, which refines the Doppler sampling. Because the physical Doppler frequencies do not lie exactly on the grid, the channel energy spreads across adjacent Doppler bins, which makes the estimation task more challenging, particularly when the number of pilot observations is limited. 
 Nevertheless, as shown in Fig. \ref{fig:frac_Doppler}, the proposed estimator maintains stable performance under fractional Doppler, and increasing $N_p$ significantly improves accuracy by providing additional observations for the expanded DD representation.\\
 
 Next, the impact of increasing the number of subcarriers \(M\) is examined. In OTFS, the delay resolution is given by \(\Delta\tau = 1/(M\Delta f)\); thus, increasing \(M\) refines the physical delay grid. However, the size of the inverse problem depends on the modeled DD grid \(M_{\tau}N_{\nu}\) and the pilot length \(N_p\), and does not depend directly on \(M\). As seen in Fig. \ref{fig:scaled_M}, it can be concluded that increasing $M$ from 32 to 64 with fixed $M_{\tau}$ results in a small variation in estimation accuracy. Furthermore, when $M_{\tau}$ is moderately increased (while still satisfying $M_{\tau} \ll M$), the performance is effectively restored.
 Therefore, increasing \(M\) enhances delay resolution without increasing computational complexity since $M_{\tau}<< M$, demonstrating that the proposed method remains scalable without additional computational overhead.

\begin{figure*}[!h]
  \centering
  \begin{subfigure}[b]{0.48\linewidth}
    \centering
    \includegraphics[height=7.5cm]{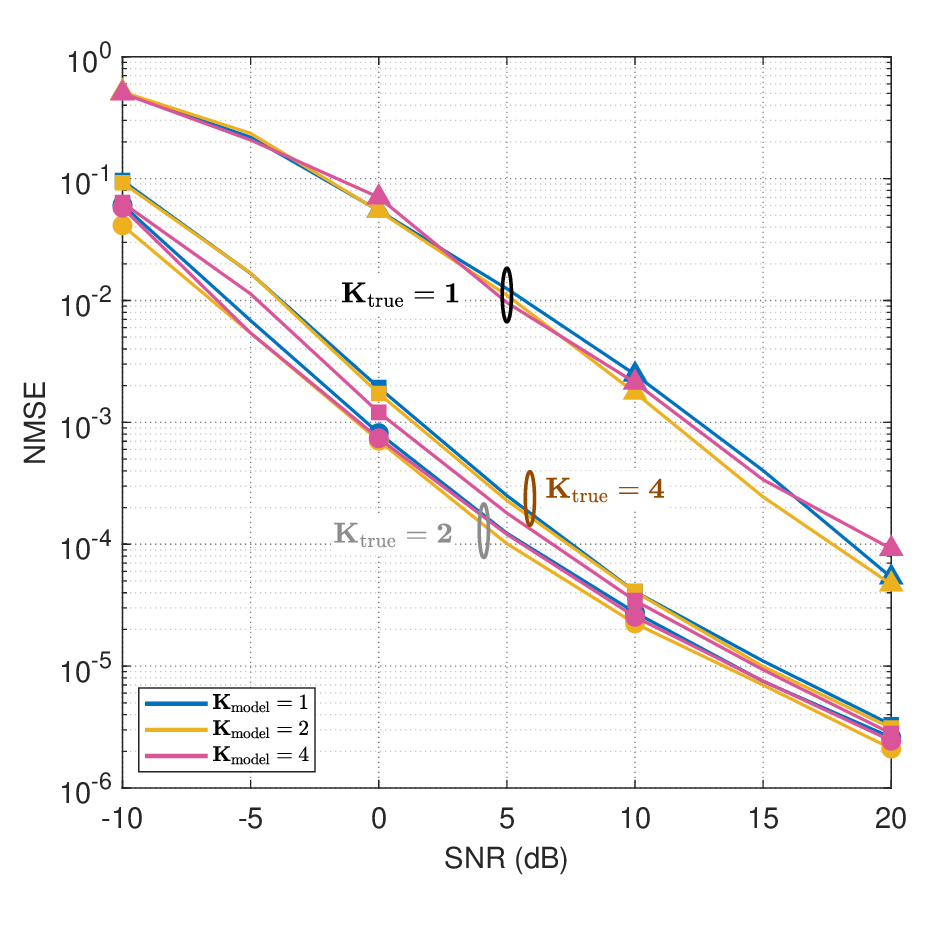}
    \caption{NMSE vs SNR across different $K$.}
    \label{fig:K}
  \end{subfigure}\hfill
  \begin{subfigure}[b]{0.49\linewidth}
    \centering
    \includegraphics[height=7.5cm]{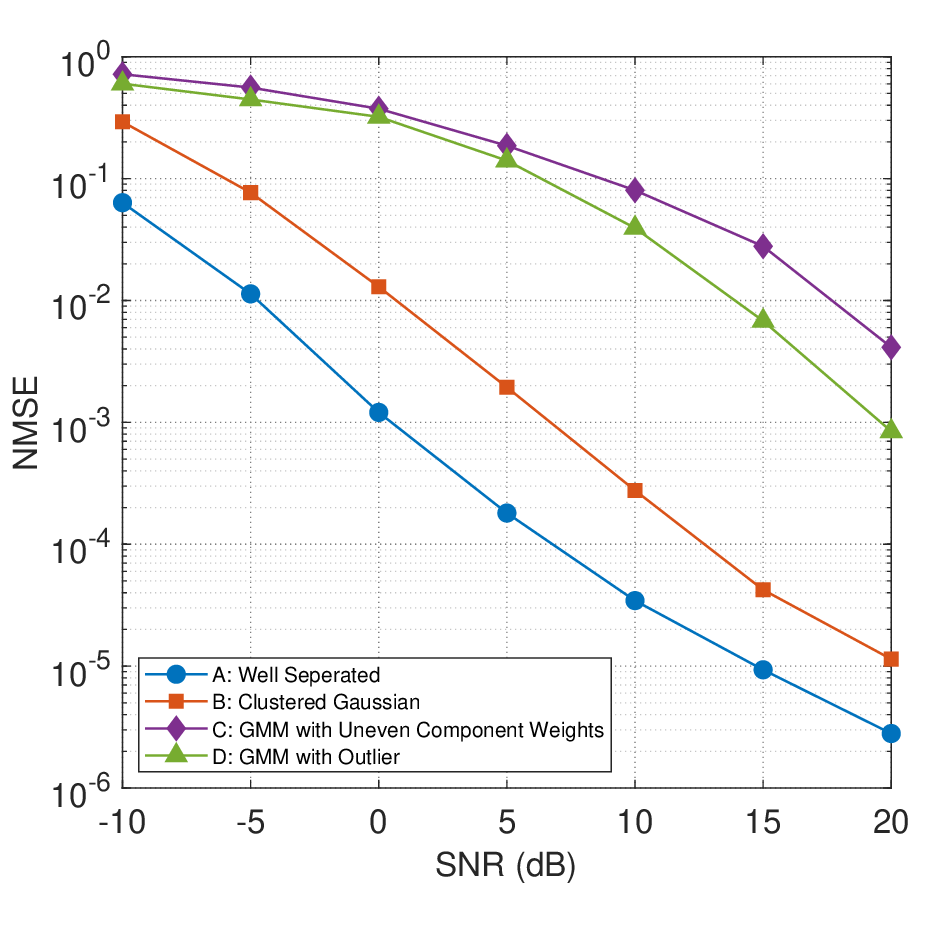}
    \caption{ NMSE vs SNR for GMM component separation cases.}
    \label{fig:gmm_means}
  \end{subfigure}

  \vspace{2mm}
  \caption{Analysis of the proposed GMM-SBL algorithm with clustered channel.}
  \label{fig:K_gmm}
\end{figure*}

\subsection{Impact of Mixture Order Selection}

To study the effect of model-order mismatch, channels were generated with 
$K_{\text{true}}=1,2,4$ and the GMM-SBL method was evaluated for 
$K_{\text{model}}\in\{1,2,4\}$. These values were selected to cover 
$K=1$, which corresponds to a single Gaussian prior (commonly used in channel modeling). Furthermore,
$K=2$ represents the simplest Gaussian mixture case, and 
$K=4$ allows the examination of a higher-order mixtures. The mixture weights $(\rho_{k})$, mean $(\mu_k)$ and variances $(\sigma_k^{2})$ are chosen to yield balanced component contributions along with well-separated clusters, thereby identifying the number of clusters $K$ to be the central objective. As observed in Fig. \ref{fig:K}, $K_{\text{true}}=1$, $K_{\text{model}}=1$ corresponds to the standard single Gaussian prior, which is equivalent to the baseline assumption in SBL.

In particular, when the underlying channel follows a two-component mixture ($K_{\text{true}} = 2$), the proposed GMM-SBL with $K_{\text{model}} = 2$ achieves approximately $13\%$ NMSE reduction at 0 dB and nearly $18\%$ at 5 dB compared to conventional single Gaussian SBL ($K_{\text{model}} = 1$). For more heterogeneous channels ($K_{\text{true}} = 4$), the improvement becomes more pronounced; at 0 dB, the NMSE reduction exceeds $35\%$ relative to the single Gaussian SBL baseline. When the true channel distribution is unimodal ($K_{\text{true}} = 1$), the single Gaussian configuration ($K_{\text{model}} = 1$) achieves the lowest NMSE among the evaluated models, while higher order mixture configurations result in only minor performance variation. These results demonstrate that the performance gain of the proposed GMM-SBL framework arises from improved statistical alignment between the assumed prior and the underlying channel distribution.
 
 The improved performance for $K_{\text{model}}=2$, can be attributed to the enhanced flexibility to capture variations in the channel coefficient distribution that are not well represented by a single Gaussian. By contrast, $K_{\text{model}}=4$ leads to a slight degradation due to over-parameterisation. For $K_{\text{true}}=2$, the matched case with $K_{\text{model}}=2$ yields the best NMSE, since the estimator correctly reflects the underlying two-component Gaussian mixture. Here, $K_{\text{model}}=1$ underfits by imposing a single-Gaussian approximation, while $K_{\text{model}}=4$ redistributes weights across unnecessary components and introduces estimation noise. For $K_{\text{true}}=4$, the estimator with $K_{\text{model}}=4$ best aligns with the four-component prior, while $K_{\text{model}}=2$ provides a reasonable approximation by grouping components, and $K_{\text{model}}=1$ shows consistently higher NMSE, especially at low SNR. Overall, fixing $K=2$ represents a robust and computationally efficient compromise; it exactly captures the two-component case, while remaining sufficiently expressive for unimodal and higher-order mixtures, thus generalising well across practical sparse CSI estimation scenarios in OTFS systems, making $K=2$ the most suitable choice for our work.

\subsection{Effect of Mixture Configurations}
To explore the full potential of the GMM-SBL algorithm, we further tested diverse scenarios having channel distributions following different  Gaussian mixtures, as described in Table \ref{tab:gmm_cases_wide}. Specifically, four representative cases having $K=4$ are examined to span well-separated, partially overlapping, variance-differentiated, and outlier-dominated conditions, by suitably choosing their mean and variance.

As depicted in Fig. \ref{fig:gmm_means}, in Case A, the best performance is achieved, since well seperated means and small variances ensure minimal overlap, enabling the estimator to identify the correct channel taps with high precision. In Case B, a moderate degradation is observed because, closely spaced means cause interference between taps within each pair. Case C is the most challenging one, having the worst performance, because identical means and differing variances produce a distribution that resembles a single dominant cluster at low SNR, forcing the estimator to purely rely on variance-based separation. Finally, Case D highlights noise sensitivity with a loss in performance since infrequent, high-amplitude channel taps are difficult to detect under noisy conditions, but become recoverable at higher SNR.

\begin{table}[!b]
\footnotesize
\renewcommand{\arraystretch}{0.8}
\caption{Parameterisation of GMMs for Channel Generation}
\centering
\label{tab:gmm_cases_wide}

\begin{tabular}{|c|l|l|}
\hline
\multicolumn{1}{|c|}{\textbf{Case}} & 
\multicolumn{1}{c|}{\textbf{GMM Description}} & 
\multicolumn{1}{c|}{\textbf{Mixture Weights} $\boldsymbol{\rho_k}$} \\ \hline

A & Well-Separated & $[0.25,\;0.25,\;0.25,\;0.25]$ \\ \hline

B & Clustered Gaussian & $[0.25,\;0.25,\;0.25,\;0.25]$ \\ \hline

C & GMM with Uneven Component & $[0.4,\;0.1,\;0.4,\;0.1]$ \\ 
  & Weights & \\ \hline

D & Gaussian Mixture with Outlier & $[0.7,\;0.15,\;0.1,\;0.05]$ \\ \hline

\end{tabular}
\end{table}

\subsection{Computational Complexity and Pilot Overhead}

 It can be observed from Table~\ref{tab:csgmm_cc} that the per-iteration complexity scales linearly with the maximum delay and Doppler support $(M_\tau, N_\nu)$
 and with the mixture order $K$, while the cubic term arises from the inversion of covariance matrices of dimension $N_p \times N_p$. Therefore, the computational burden depends primarily on the pilot dimension $N_p$ and the sparse DD grid parameters $(M_\tau, N_\nu)$, rather than on the full OTFS frame size $(M,N)$. Since typically $M_{\tau} \ll M$ and $N_{\nu} \ll N$, the effective inverse problem dimension remains significantly smaller than the full OTFS grid. Additionally, for the configurations considered in this work ($N_p = 80$, $K = 2$), accurate estimation performance is achieved without requiring excessive pilot dimensions or high mixture orders, resulting in moderate computational complexity.
 
 Further, computational cost in the work is a justified trade-off, because GMM-SBL learns a data-driven, multimodal prior that captures complex channel structures. Consequently, GMM-SBL achieves significantly lower NMSE and BER, providing robust performance even in low-SNR regimes.

\begin{table}[!t]
\scriptsize
\renewcommand{\arraystretch}{1.15}
\centering
\caption{Computational Complexity of GMM--SBL}
\label{tab:csgmm_cc}
\begin{tabular}{|l|c|c|}
\hline
\textbf{Operation} & \textbf{Complex multiplications} & \textbf{Complex additions} \\ \hline

\makecell[l]{\footnotesize 
Formation of\\ $\mathbf{A}_k =$\\ $\sigma^2\mathbf{I} 
+ \boldsymbol{\Omega}\boldsymbol{\Gamma}_k\boldsymbol{\Omega}^H$} 
& $K\!\left( N_p^2 M_\tau N_\nu + N_p^3 \right)$ 
& $K\!\left( N_p^2 M_\tau N_\nu + N_p^3 \right)$ \\ \hline

\makecell[l]{Update of $\mathbf{u}_{i,k}$\\ 
$\mathbf{u}_{i,k} = \mathbf{A}_k^{-1}\mathbf{r}_{p,i}$} 
& $K L N_p^2$ 
& $K L N_p (N_p - 1)$ \\ \hline

\makecell[l]{Posterior mean  \\ 
$\boldsymbol{\mu}_{i,k} = \boldsymbol{\Gamma}_k \boldsymbol{\Omega}^H \mathbf{u}_{i,k}$} 
& $K L M_\tau N_\nu (N_p + 1)$ 
& $K L M_\tau N_\nu (N_p - 1)$ \\ \hline

\makecell[l]{Diagonal covariance \\ 
$\boldsymbol{\Sigma}_k = \boldsymbol{\Gamma}_k -$ \\
$\boldsymbol{\Gamma}_k \boldsymbol{\Omega}^H \mathbf{A}_k^{-1} \boldsymbol{\Omega}\boldsymbol{\Gamma}_k$} 
& $K M_\tau N_\nu N_p^2$ 
& $K M_\tau N_\nu N_p^2$ \\ \hline

\makecell[l]{Hyperparameter \\ updates \\ 
$(\boldsymbol{\Gamma}_k,\, \rho_k)$} 
& $K M_\tau N_\nu (L + 1)$ 
& $K (L - 1)(1 + M_\tau N_\nu)$ \\ \hline

\end{tabular}
\end{table}

The proposed method inserts pilots directly in the TD, giving a pilot overhead of $\alpha_{\text{t}} = \tfrac{N_p}{MN+N_p}$. With the parameters of Table \ref{sub6parameters}, this evaluates to about $0.0725$. For context and fair comparison to the proposed method, both DD and TF domain pilot schemes from existing SBL-based CSI estimation are considered. A  DD domain pilot scheme without guard symbols and with approximate overhead of $\alpha_{\text{DD}} \approx \tfrac{(4N_\nu+1)M_\tau}{MN}$ is discussed in \cite{zhao2020sparse}. Although efficient, this expression depends on $(N_\nu, M_\tau)$, therefore leading to enhanced pilot overhead in highly dispersive channels. A TF domain pilot scheme proposed in \cite{srivastava2021bayesian} gives $\alpha_{\text{TF}} = \tfrac{N_p}{N+N_p}$. Although this scheme avoids dependency on $(N_\nu, M_\tau)$, it requires $MN_p$ pilot symbols and involves additional linear transforms and structured matrix operations for pilots, leading to increased receiver and transmitter complexity. By contrast, the proposed TD scheme transmits only $N_p$ raw pilot symbols, avoids any DD guard interval, and yields a direct input–output relationship for sparse  DD domain CSI recovery, thereby providing a simpler transceiver structure and a lower effective pilot overhead.

\section{Conclusions}
\label{Conclusion} 
To enhance the channel estimation performance of OTFS systems, a GMM-SBL framework was proposed, which exploits the essential  DD domain sparsity, while significantly improving statistical modeling using a mixture prior. A unified EM algorithm jointly refines mixture weights and diagonal variances directly from the raw TD pilot observations. The framework incorporates key innovations such as a hierarchical prior structure that readily adapts to both simple and complex channel conditions through learned mixture components. It also provides a practical implementation using rectangular pulse shaping with flexible pilot placement, which eliminates DD guard interval requirements, substantially reducing overhead while preserving spectral efficiency. The estimated CSI is subsequently utilised in an LMMSE detector for reliable data detection. Moreover, our theoretical analysis has established sparsity guarantees for the proposed mixture priors, while the Oracle-MMSE and BCRLB benchmarks derived provide rigorous performance references.
Comprehensive simulations across diverse channel conditions, including single-Gaussian and multi-component mixtures with varying parameters, demonstrate consistent and substantial improvements in both estimation accuracy and detection reliability over the state-of-the-art sparse methods.

\appendices

\section{Proof of Theorem 1}
\label{appendix a}

For each component $k$, the Gaussian prior 
$p(\mathbf{h} \mid z{=}k) = \mathcal{CN}(\mathbf{0}, \boldsymbol{\Gamma}_k)$ 
has zero mean along with diagonal covariance. For each such component, there exists a constant $C> 0$ such that:
\begin{equation}
\mathcal{CN}(\mathbf{0}, \boldsymbol{\Gamma}_k) 
\;\leq\; C\cdot \prod_{r=1}^{M_\tau N_\nu} \frac{1}{|h_r|^{2}}.
\end{equation}
The mixture prior is a convex combination of these components:
\begin{align}
p(\mathbf{h}) =& \sum_{k=1}^{K} \rho_k~ p(\mathbf{h} \mid z{=}k) \leq \left( \sum_{k=1}^{K} \rho_k C_k \right)
\cdot \prod_{r=1}^{M_\tau N_\nu} \frac{1}{|h_r|^{2}}\nonumber \\ =& C \cdot \prod_{r=1}^{M_\tau N_\nu} \frac{1}{|h_r|^{2}},~~~~ \text{where}~~~~ C = \sum_{k=1}^{K} \rho_k C_k.
\end{align}

\section{Proof of Theorem 2}
\label{appendix b}

\renewcommand{\qedsymbol}{} 

Let $\mathbf{R'} = \{ \mathbf{r}_{p,i} \}_{i=1}^L$ denote the observed snapshots. For \(k\in\{1,\dots,K\}\) assume component priors
\[
p(\mathbf{h}\mid k)=\mathcal{CN}(\mathbf{0},\boldsymbol{\Gamma}_k),\qquad \boldsymbol{\Gamma}_k=\mathrm{diag}(\gamma_{k,1},\ldots,\gamma_{k,M_\tau N_\nu}),
\]
and mixture weights \(\rho_k\), with \(\sum_{k=1}^{K}\rho_k=1\).
Define responsibilities at EM iteration \(t\) by
\begin{equation}\label{eq:resp_def}
\pi_{i,k}^{(t)} \;:=\; p^{(t)}(k\mid \mathbf{r}_{p,i})
= \frac{\rho_k^{(t)}\,p(\mathbf{r}_{p,i}\mid k;\boldsymbol{\Gamma}_k^{(t)})}{\sum_{\ell=1}^K \rho_\ell^{(t)}\,p(\mathbf{r}_{p,i}\mid \ell;\boldsymbol{\Gamma}_\ell^{(t)})},
\end{equation}
where the Gaussian marginal is
\begin{equation}\label{eq:marginal}
p(\mathbf{r}_{p,i}\mid k) \;=\; \mathcal{CN}\big(\boldsymbol{0},\; \sigma^{2}\mathbf{I} + \boldsymbol{\Omega}\boldsymbol{\Gamma}_k\boldsymbol{\Omega}^{H}\big).
\end{equation}
Let the conditional posterior under component \(k\) for snapshot \(i\) be
\(\;p(\mathbf{h}\mid \mathbf{r}_{p,i},k)=\mathcal{CN}(\boldsymbol{\mu}_{i,k},\boldsymbol{\Sigma}_{k})\;\) where \(\boldsymbol{\mu}_{i,k},\boldsymbol{\Sigma}_{k}\) represents the a-posteriori mean and covariance, after substituting \(\boldsymbol{\Gamma}\leftarrow\boldsymbol{\Gamma}_k\), the M-step yields

\begin{subequations}
\begin{align}
\gamma_{k,r}^{(t+1)}
&= 
\frac{\displaystyle\sum_{i=1}^{L}\pi_{i,k}^{(t)}
\big(|\mu_{i,k,r}|^{2} + [\boldsymbol{\Sigma}_{k}]_{rr}\big)}
{\displaystyle\sum_{i=1}^{L}\pi_{i,k}^{(t)}}, \label{eq:gamma_rho_updates_a}\\
\rho_k^{(t+1)} 
&= 
\frac{1}{L}\sum_{i=1}^{L}\pi_{i,k}^{(t)}.\label{eq:gamma_rho_updates_b}
\end{align}
\label{eq:gamma_rho_updates}
\end{subequations}
To prove this, let the complete data for the snapshot \(i\) be \(p(\mathbf{r}_{p,i},\mathbf{h},k)=p(\mathbf{r}_{p,i}\mid \mathbf{h})\,p(\mathbf{h}\mid k)\,\rho_k\).
Only the prior \(p(\mathbf{h}\mid k)\) and the mixing weight \(\rho_k\) depend on \(\boldsymbol{\Gamma}_k\) and $\rho$, respectively.
Thus, the EM functional at iteration \(t\) is

\begin{align}
\mathcal{Q}^{(t)}&(\{\boldsymbol{\Gamma}_k\},\rho)
= \sum_{i=1}^{L} \sum_{k=1}^{K} \pi_{i,k}^{(t)} \Bigg(
   - \sum_{r=1}^{M_\tau N_\nu} \log \gamma_{k,r} \nonumber \\
&\quad - \sum_{r=1}^{M_\tau N_\nu} 
   \frac{\mathbb{E}_{\mathbf{h}\mid \mathbf{r}_{p,i},k}[\,|h_r|^{2}\,]}{\gamma_{k,r}}
   + \log \rho_k \Bigg) + C ,\label{eq:Q_def}
\end{align}

where \(C\) is independent of \(\boldsymbol{\Gamma}_k\ \text{and} ~\rho\). By using the identity
\begin{equation}\label{eq:post_moment}
\mathbb{E}_{\mathbf{h}\mid \mathbf{r}_{p,i},k}\big[|h_r|^{2}\big]
= |\mu_{i,k,r}|^{2} + [\boldsymbol{\Sigma}_{k}]_{rr},
\end{equation}
we isolate the scalar function of \(\gamma_{k,r}\)
\[
Q_{k,r}(\gamma_{k,r}) = -\sum_{i=1}^{L}\pi_{i,k}^{(t)}\Big(\log\gamma_{k,r}
+ \frac{|\mu_{i,k,r}|^{2}+[\boldsymbol{\Sigma}_{k}]_{rr}}{\gamma_{k,r}}\Big).
\]
We differentiate \(Q_{k,r}\) with respect to \(\gamma_{k,r}\) and set it to zero:
\[
\frac{\partial Q_{k,r}}{\partial\gamma_{k,r}}
= -\sum_{i=1}^{L}\pi_{i,k}^{(t)}\Big(\frac{1}{\gamma_{k,r}}
- \frac{|\mu_{i,k,r}|^{2}+[\boldsymbol{\Sigma}_{k}]_{rr}}{\gamma_{k,r}^{2}}\Big)=0.
\]
Upon multiplying both sides by \(\gamma_{k,r}^{2}\) and rearranging, we obtain
\[
\gamma_{k,r}^{(t+1)} \sum_{i=1}^{L}\pi_{i,k}^{(t)}
= \sum_{i=1}^{L}\pi_{i,k}^{(t)}\big(|\mu_{i,k,r}|^{2}+[\boldsymbol{\Sigma}_{k}]_{rr}\big),
\]
which yields \eqref{eq:gamma_rho_updates_a}.

To update \(\rho\) retain only the \(\rho\)-dependent part of \(\mathcal{Q}^{(t)}\) in \eqref{eq:Q_def}:
\[
\mathcal{Q}^{(t)}_{\rho} = \sum_{i=1}^{L}\sum_{k=1}^{K}\pi_{i,k}^{(t)}\log\rho_k.
\]
Then introduce the Lagrange multiplier \(\beta\) for \(\sum_{k=1}^{K}\rho_k=1\) and form
\(\Lambda(\rho,\beta)=\mathcal{Q}^{(t)}_{\rho} + \beta\big(\sum_{k=1}^{K}\rho_k-1\big)\).
Differentiating with respect to \(\rho_k\) yields

\[
\frac{\partial\Lambda}{\partial\rho_k}
= \frac{\sum_{i=1}^{L}\pi_{i,k}^{(t)}}{\rho_k} + \beta = 0
\quad\Longrightarrow\quad
\rho_k = -\frac{1}{\beta}\sum_{i=1}^{L}\pi_{i,k}^{(t)}.
\]

Normalization \(\sum_{k=1}^{K}\rho_k=1\) gives
\[
-\frac{1}{\beta}\sum_{k=1}^{K}\sum_{i=1}^{L}\pi_{i,k}^{(t)} = 1.
\]
Since \(\sum_{k=1}^{K}\pi_{i,k}^{(t)}=1\) for every \(i\), the double sum equals \(L\),
hence we have \(-1/\beta=1/L\) and therefore \(\rho_k^{(t+1)}=\tfrac{1}{L}\sum_{i=1}^{L}\pi_{i,k}^{(t)}\),
which is \eqref{eq:gamma_rho_updates_b}. 

 \medskip

\section{BCRLB under GMM Prior}
\label{appendix c}

Let the dictionary be \(\boldsymbol{\Omega}\in\mathbb{C}^{N_p\times (M_\tau N_\nu)}\) and the DD domain channel be
\(\mathbf{h}\in\mathbb{C}^{M_\tau N_\nu\times 1}\). With \(L\) TD pilot snapshots
\(\{\mathbf{r}_{p,i}\}_{i=1}^L\) under additive noise
\(\boldsymbol{\eta}^{(i)}\sim\mathcal{CN}(\mathbf{0},\sigma^2\mathbf{I})\),
the Bayesian Fisher information matrix (FIM) decomposes as in~\cite{crafts2024bayesian}
\begin{equation}
\mathbf{J} \;=\; \mathbf{J}_{\mathrm{data}} + \mathbf{J}_{\mathrm{prior}},
\label{eq:J_total}
\end{equation}
where, $\mathbf{J}_{\mathrm{data}} = \tfrac{L}{\sigma^2}\boldsymbol{\Omega}^H\boldsymbol{\Omega}$ 
represents the data Fisher information matrix, while 
$\mathbf{J}_{\mathrm{prior}}$ denotes the contribution from the prior, which can be derived as follows.

\medskip
Assume a \(K\)-component complex Gaussian mixture prior
\begin{equation}
p(\mathbf{h}) \;=\; \sum_{k=1}^K \rho_k\,\mathcal{CN}(\boldsymbol{\mu}_k,\boldsymbol{\Gamma}_k),
\quad \rho_k\!\ge\!0,~ \sum_{k=1}^K \rho_k\!=\!1,
\label{eq:gmm_prior}
\end{equation}
where \(\rho_k\) are the constant mixture weights. Let us define the local weights as
\begin{equation}
w_k(\mathbf{h}) \;=\; 
\frac{\rho_k\,\mathcal{CN}(\boldsymbol{\mu}_k,\boldsymbol{\Gamma}_k)}{p(\mathbf{h})},
\qquad \sum_{k=1}^K w_k(\mathbf{h}) = 1.
\label{eq:wk_def}
\end{equation}
Then differentiation of the mixture log-density yields the Hessian identity:
\begin{align}
&\nabla_{\mathbf{h}}^2 [\log p(\mathbf{h})]
= \sum_{k=1}^K w_k(\mathbf{h})
\Big[
-\boldsymbol{\Gamma}_k^{-1} \nonumber\\
&\hspace{-1.5em} \quad\;\; + \boldsymbol{\Gamma}_k^{-1}
(\mathbf{h}-\boldsymbol{\mu}_k)(\mathbf{h}-\boldsymbol{\mu}_k)^H
\boldsymbol{\Gamma}_k^{-1}
\Big] \quad - \mathbf{b}(\mathbf{h})\mathbf{b}(\mathbf{h})^H .
\label{eq:hessian_identity}
\end{align}
where we have 
\begin{equation}
\mathbf{b}(\mathbf{h}) \;=\; 
\sum_{k=1}^K w_k(\mathbf{h})
\big[-\boldsymbol{\Gamma}_k^{-1}(\mathbf{h}-\boldsymbol{\mu}_k)\big].
\label{eq:b_vector}
\end{equation}
Taking the negative expectation with respect to \(p(\mathbf{h})\) yields the prior Fisher information matrix as follows:
\begin{align}
\mathbf{J}_{\mathrm{prior}}
&= -\mathbb{E}_{\mathbf{h}}\!\big[\nabla_{\mathbf{h}}^2\log p(\mathbf{h})\big] = \mathbb{E}_{\mathbf{h}}\!\Big[\sum_{k=1}^K w_k(\mathbf{h})\,\boldsymbol{\Gamma}_k^{-1}\Big] \nonumber\\
&\quad - \mathbb{E}_{\mathbf{h}}\!\Big[\sum_{k=1}^K w_k(\mathbf{h})\,\boldsymbol{\Gamma}_k^{-1}
(\mathbf{h}-\boldsymbol{\mu}_k)(\mathbf{h}-\boldsymbol{\mu}_k)^H
\boldsymbol{\Gamma}_k^{-1}\Big] \nonumber\\
&\quad + \mathbb{E}_{\mathbf{h}}\!\big[\mathbf{b}(\mathbf{h})\mathbf{b}(\mathbf{h})^H\big].\label{eq:J_prior}
\end{align}

For \(K=1\), \eqref{eq:J_prior} reduces to \(\mathbf{J}_{\mathrm{prior}}=\boldsymbol{\Gamma}_1^{-1}\).
The total Bayesian information matrix is then obtained by substituting \eqref{eq:J_prior} into \eqref{eq:J_total}, giving
\begin{equation}
\mathbf{J}
\;=\;
\tfrac{L}{\sigma^2}\boldsymbol{\Omega}^H\boldsymbol{\Omega}
+\mathbf{J}_{\mathrm{prior}}.
\label{eq:J_final}
\end{equation}
Then the BCRLB of the average MSE is expressed as
\begin{equation}
\mathrm{MSE}(\widehat{\mathbf{h}}) \;\ge\;
\mathrm{tr}\!\big(\mathbf{J}^{-1}\big)
\;=\;
\mathrm{tr}\!\Big(
\big[
\tfrac{L}{\sigma^2}\boldsymbol{\Omega}^H\boldsymbol{\Omega}
+ \mathbf{J}_{\mathrm{prior}}
\big]^{-1}
\Big).
\label{eq:BCRLB}
\end{equation}

For \(K>1\), the expectations in \eqref{eq:J_prior} do not admit closed-form solutions and are therefore
approximated using Monte–Carlo simulation. In practice, \(L\) i.i.d. draws
\(\mathbf{h}^{(i)} \sim p(\mathbf{h})\) are generated, and the local weights
\(w_k(\mathbf{h}^{(i)})\) are evaluated for each sample, and the resultant
per-sample contributions are averaged to form an empirical prior Fisher
information matrix, \(\widehat{\mathbf{J}}_{\mathrm{prior}}\), which is then substituted
into \eqref{eq:BCRLB} to obtain the BCRLB.

\bibliographystyle{IEEEtran}
\bibliography{bibliography.bib}

\end{document}